\documentclass[prb,twocolumn,showpacs]{revtex4-1}%
\usepackage{amsfonts}
\usepackage{amsmath}
\usepackage{amssymb}
\usepackage{graphicx}
\usepackage{txfonts}%
\begin{document}
\title{General Green's function formalism for layered systems:\ Wave function approach}
\author{Shu-Hui Zhang$^{1}$}
\author{Wen Yang$^{1}$}
\email{wenyang@csrc.ac.cn}
\author{Kai Chang$^{2,3}$}
\email{kchang@semi.ac.cn}
\affiliation{$^{1}$Beijing Computational Science Research Center, Beijing 100193, China}
\affiliation{$^{2}$SKLSM, Institute of Semiconductors, Chinese Academy of Sciences, P.O.
Box 912, Beijing 100083, China}
\affiliation{$^{3}$Synergetic Innovation Center of Quantum Information and Quantum Physics,
University of Science and Technology of China, Hefei, Anhui 230026, China}

\begin{abstract}
The single-particle Green's function (GF) of mesoscopic structures plays a
central role in mesoscopic quantum transport. The recursive GF technique is a
standard tool to compute this quantity numerically, but it lacks physical
transparency and is limited to relatively small systems. Here we present a
numerically efficient and physically transparent GF formalism for a general
layered structure. In contrast to the recursive GF that directly calculates
the GF through the Dyson equations, our approach converts the calculation of
the GF to the generation and subsequent propagation of a scattering wave
function emanating from a local excitation. This viewpoint not only allows
us to reproduce existing results in a concise and physically intuitive manner,
but also provides analytical expressions of the GF in terms of a generalized
scattering matrix. This identifies the contributions from each individual
scattering channel to the GF and hence allows this information to be
extracted quantitatively from dual-probe STM experiments. The simplicity and
physical transparency of the formalism further allows us to treat the multiple
reflection analytically and derive an analytical rule to construct the GF of a
general layered system. This could significantly reduce the computational time
and enable quantum transport calculations for large samples. We apply this
formalism to perform both analytical analysis and numerical simulation for the two-dimensional conductance map of a realistic
graphene \textit{p-n} junction. The results demonstrate the possibility of observing
the spatially-resolved interference pattern caused by negative refraction and
further reveal a few interesting features, such as the distance-independent
conductance and its quadratic dependence on the carrier concentration, as
opposed to the linear dependence in uniform graphene.

\end{abstract}

\maketitle

\section{Introduction}

The single-particle retarded Green's function (GF) is a key tool to calculate
local and transport properties in mesoscopic systems
\cite{DattaBook1995,FerryBook1997}, such as conductance, shot noise
\cite{BlanterPhysRep2000}, local density of states, and local currents
\cite{CrestiPRB2003,MetalidisPRB2005}. In the most popular scheme in which a
scatterer is connected to two (or more) semi-infinite ballistic leads, the
Landauer-B\"{u}ttiker formula
\cite{LandauerJRD1957,LandauerPM1970,BuettikerPRB1985} expresses the
conductance $\sigma=(e^{2}/h)T(E_{F})$ in terms of the transmission
probability $T(E_{F})$ across the scatterer on the Fermi surface. Typically, the electronic
structure and transport properties of a mesoscopic system are described by a
lattice model with a localized basis in real space, e.g., discretization of the
continuous model \cite{KhomyakovPRB2004}, empirical tight-binding
\cite{HarrisonBook1989}, or first-principles density-functional theory with a
localized basis set \cite{KudrnovskyPRB1994,ZellerPRB1995}. Then $T(E_{F})$ is
constructed from the lattice GF $G(E_{F})$ across the scatterer through either an expression
derived by Caroli \textit{et al.} \cite{CaroliJPC1971} or the Fisher-Lee
relations \cite{FisherPRB1981,StoneIJRD1988,BarangerPRB1989,SolsAP1992} that
express the scattering matrix $S(E_{F})$ of the scatterer in terms of the
lattice GF $G(E_{F})$.

The recursive GF method (RGF) is a standard tool to compute the lattice GF of
a scatterer connected to multiple leads \cite{FerryBook1997,LewenkopfJCE2013}.
This method is reliable, computationally efficient, and allows for a parallel
implementation \cite{DrouvelisJCP2006}. It was pioneered by Thouless and
Kirkpatrick \cite{ThoulessJPC1981} and by Lee and Fisher \cite{LeePRL1981}.
Then MacKinnon presented a \textquotedblleft slice\textquotedblright%
\ formulation for a general layered system, which is the form most used
nowadays \cite{MacKinnonZPB1985}. Variations of the method were also introduced
to treat multiple leads \cite{BarangerPRB1991}, arbitrary geometries
\cite{KazymyrenkoPRB2008,WimmerThesis2009}, and local scatterers inside an
infinite periodic system \cite{ZhengJCP2010,SettnesPRB2015}. With the
development of many numerical algorithms, such as fast recursive or iterative
schemes
\cite{MacKinnonPRL1981,SoukoulisPRB1982,MacKinnonZPB1983,SchweitzerJPC1984,SanchoJPFMP1984,SanchoJPF1985,AokiPRL1985,AndoPRB1989,SolsJAP1989,AndoPRB1990,GodfrinJPCM1991,NardelliPRB1999}
and closed-form solutions
\cite{UmerskiPRB1997,SanvitoPRB1999,Garcia-MolinerSS1994,VasseurSSR2004,RochaPRB2004}%
, the development of RGF has culminated in many packages with different focus
\cite{BrandbygePRB2002,RochaPRB2006,BirnerIEEE2007,OzakiPRB2010,FonsecaJCE2013}
and is applicable to an arbitrary lattice Hamiltonian \cite{GrothNJP2014}. However,
these techniques and our knowledge about the lattice GF still suffer from two
drawbacks/limitations.\begin{figure}[t]
\includegraphics[width=\columnwidth,clip]{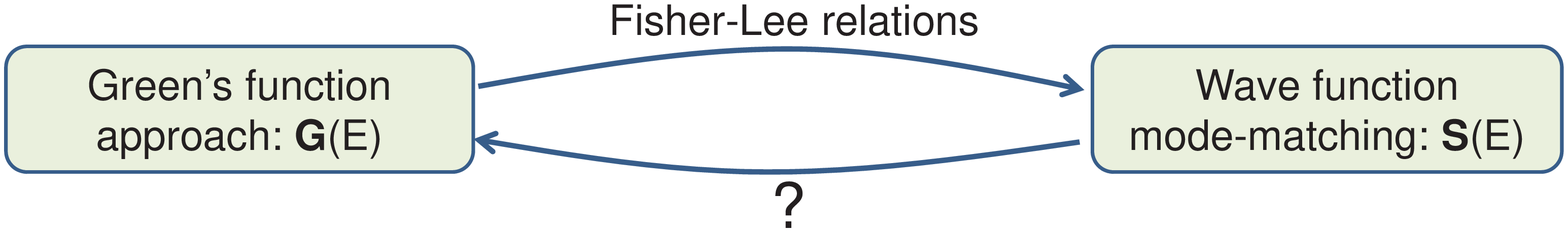}
\caption{Connection
between two widely used approaches to mesoscopic quantum transport: the GF
approach and wave function mode matching approach. The former calculates the
GF $G(E)$, while the latter calculates the scattering matrix
$S(E)$. The Fisher-Lee relations allow $S(E)$ to be
constructed from $G(E)$, while the inverse relation remains absent
for a general lattice model.}%
\label{G_GFVSMODEMATCH}%
\end{figure}

First, there are two widely used approaches in mesoscopic quantum
transport:\ the GF approach \cite{DattaBook1995,Datta2000} that computes the
GF $G(E)$ and the wave function mode matching approach
\cite{AndoPRB1991,NikolicPRB1994,XiaPRB2006} that computes the unitary
scattering matrix $S(E)$. However, the connection between the GF $G(E)$ and
$S(E)$ and hence the connection between these two approaches remain
incomplete. It is well known that $S(E)$ can be constructed from $G(E)$
through the Fisher-Lee relations (see Fig. \ref{G_GFVSMODEMATCH}), as first
derived by Fisher and Lee \cite{FisherPRB1981} for a single parabolic band and
two-terminal structures and later generalized to multiple leads
\cite{StoneIJRD1988,BarangerPRB1989,SolsAP1992} and arbitrary lattice models
\cite{SanvitoPRB1999,KhomyakovPRB2005,WimmerThesis2009}. However, the inverse
of this relation is nontrivial \footnote{Since $G(E)$ describes the scattering
of both traveling and \textit{evanescent} waves, while $S(E)$ only describes
the scattering of traveling waves, $G(E)$ and $S(E)$ are equivalent only in
the asymptotic regions of the lead, i.e., sufficiently far way from the
scatterer such that the evanescent waves vanish completely.}:\ explicit
expressions of $G(E)$ in terms of $S(E)$ were derived
\cite{StoneIJRD1988,BarangerPRB1989,SolsAP1992} only for a single parabolic
band and in regions far from the scatterer. Generalization of this result to a
general lattice model and over other regions would not only complete the
equivalence \cite{KhomyakovPRB2005} between the GF approach and the wave
function mode matching approach (Fig. \ref{G_GFVSMODEMATCH}), but also
provides important tools to analyze the multi-probe scanning tunneling
microscopy (STM), which has been applied to characterize a wide range of
systems (see Refs. \onlinecite{NakayamaAM2012,LiAFM2013} for recent reviews)
in the past few years, including nanowires
\cite{KuboAPL2006,CherepanovRSI2012,QinRSI2012}, carbon nanotubes
\cite{WatanabeAPL2001}, graphene nanoribbons
\cite{BaringhausNature2014,BaringhausPRL2016}, monolayer and bilayer graphene
\cite{SutterNatMater2008,JiNatMater2012,EderNanoLett2013}, and grain
boundaries in graphene \cite{ClarkAcsNano2013,ClarkPRX2014} and copper
\cite{KimNanoLett2010}. With one STM probe at $\mathbf{R}_{1}$ and the other
STM\ probe at $\mathbf{R}_{2}$, the Landauer-B\"{u}ttiker formula expresses
the conductance between the STM probes in terms of the GF $G(\mathbf{R}%
_{2},\mathbf{R}_{1},E_{F})$, which provides spatially resolved information
about the sample; e.g., with an analytical expression for the GF of pristine
graphene \cite{ZhengJCP2010,SettnesPRB2015}, Settnes \textit{et al.
}\cite{SettnesPRL2014} were able to identify the different scattering
processes of local scatterers in graphene. However, this analysis is still
qualitative. To go one step further to extract \textit{quantitatively} the
information about the scatterers, an explicit expression of the GF in terms of
the scattering matrix is required.

Second, the time cost of RGF increases rapidly with the number of localized
basis required to subtend the sample. This imposes a computational limit when
addressing realistic experimental samples; e.g., many quantum transport
studies on graphene consider narrow graphene \textquotedblleft
nanoribbons\textquotedblright\ rather than large-area graphene. Three methods
have been proposed to lift this constraint. The modular RGF
\cite{SolsJAP1989,SolsAP1992,RotterPRB2000} is limited to electrons in a
single parabolic band and specific shape of the sample. The other two methods
essentially reduce the number of transverse bases, either by projecting the
system Hamiltonian onto a small number of transverse modes
\cite{MaaoPRB1994,ZozoulenkoPRB1996,ZozoulenkoPRB1996a} or by assuming
translational invariance \cite{LiuPRB2012} along the transverse direction.
They could significantly reduce the time cost for wide samples, but the time
cost still increases linearly with the length of the scatterer (along which
transport occurs). To study a\textit{ long} sample, a more efficient method is desirable.

The origin of the above drawbacks/limitations is probably that the RGF treats
the GF as a matrix and constructs the GF by a series of matrix recursion rules
derived from the Dyson equation. Interestingly, although the rules for
constructing the scattering matrix in terms of the GF (i.e., the Fisher-Lee
relation) are concise and physically intuitive, their rigorous derivation (in
which the GF is treated as a matrix) turns out to be rather tedious (see, e.g.,
Refs. \onlinecite{StoneIJRD1988,BarangerPRB1989,WimmerThesis2009}). This
somewhat surprising fact suggests the possible existence of a very different
way to represent and calculate the GF. This could not only enable a
straightforward physical interpretation of the final results, but also shed
light on some previous debates
\cite{NikolicPRB1994,KrsticPRB2002,KhomyakovPRB2005} on the relationship
between different calculation techniques in mesoscopic quantum transport.

In this work, we develop a numerically efficient and physically transparent GF
formalism to address the above issues in layered systems, i.e., any system
that is non-periodic along one direction, but is finite or periodic along
the other directions. This includes a wide range of physical systems, such as
interfaces and junctions, Hall bars, nanowires, multilayers, superlattices,
carbon nanotubes, and graphene nanoribbons. Compared with the RGF that
directly calculates the GF as a matrix through the Dyson equations, our
approach converts the calculation of the GF to the generation and subsequent
propagation of a scattering wave function emanating from a local excitation.
This viewpoint provides several advantages. First, the
procedures for calculating the GF $G(E)$ becomes physically transparent and
existing results from the RGF (such as the Fisher-Lee relation) can be derived in
a concise and physically intuitive manner. Second, the GF $G(E)$ can be
readily expressed in terms of a few scattering wave functions with energy $E$.
This provides an \textit{on-shell} generalization of the standard spectral expansion in
classic textbooks on quantum mechanics
\cite{SakuraiBook1994,GriffithsBook1995,CohenBook2005}, $G(\mathbf{R}%
_{2},\mathbf{R}_{1},E)=\sum_{\lambda}\langle\mathbf{R}_{1}|\Psi_{\lambda
}\rangle\langle\Psi_{\lambda}|\mathbf{R}_{2}\rangle/(E+i0^{+}-E_{\lambda})$,
which involves \textit{all} the eigenstates $\{|\Psi_{\lambda}\rangle\}$ and eigenenergies $\{E_{\lambda}\}$ of
the system. In terms of a generalized scattering matrix $\mathcal{S}(E)$ that
describes the scattering of both traveling and evanescent waves, we further
establish a one-to-one correspondence between $G(E)$ and and $\mathcal{S}(E)$ (see Fig. 1).
This identifies the contributions from each individual scattering channel
(including evanescent channels) to the GF and hence allows this information
to be extracted quantitatively from dual-probe STM. Third, the simplicity and
physical transparency of the formalism further allows us to perform an
infinite summation of the multiple reflection between different scatterers and
arrive at an analytical construction rule for the GF of a general layered
system containing an arbitrary number of scatterers. This could make the time
cost independent of the length of the sample along the transport direction and
hence significantly speed up the calculation. By further reducing the number of
bases along the transverse direction
\cite{MaaoPRB1994,ZozoulenkoPRB1996,ZozoulenkoPRB1996a,LiuPRB2012}, our
formalism enables quantum transport calculations over macroscopic distances
and on large samples.

Recently, the chiral tunneling
\cite{KleinZP1929,KatsnelsonNatPhys2006,CheianovPRB2006} and negative
refraction \cite{CheianovScience2007,ParkNanoLett2008,MoghaddamPRL2010} of
graphene \textit{p-n} junctions have received a lot of interest and the anomalous
focusing effect was observed experimentally
\cite{LeeNatPhys2015,ChenScience2016}, but previous theoretical studies are
mostly based on the low-energy continuous model, whose validity is limited to
the vicinity of the Dirac points. Here we apply our GF approach to
perform an analytical analysis and numerical simulation for the two-dimensional conductance map of dual-probe STM experiments in
a realistic graphene \textit{p-n} junction described by the tight-binding model. The
results demonstrate the possibility of observing the spatially resolved
interference pattern caused by negative refraction and further reveals some
interesting features (such as the distance-independent conductance and its
quadratic dependence on the carrier concentration, as opposed to the linear
dependence in uniform graphene) that may also be observed in dual-probe STM experiments.

This paper is organized as follows. In Sec. II, we introduce the model, review
the commonly used RGF technique, and presents the key idea of our
approach. In Sec. III, we derive the GF of an infinite system containing a
single scatterer, as well as an analytical construction rule for the GF of a
general layered system containing an arbitrary number of scatterers. In Sec.
IV, we express the GF analytically in terms of a generalized scattering matrix
or in terms of a few scattering states on the energy shell $E$. In Sec. V, we
exemplify our results in a 1D chain and then apply it to analyze and simulate the
two-dimensional conductance map of a realistic graphene \textit{p-n} junction. Finally,
a brief conclusion is given in Sec. VI.

\section{Theoretical model and key ideas}

\begin{figure}[t]
\includegraphics[width=\columnwidth,clip]{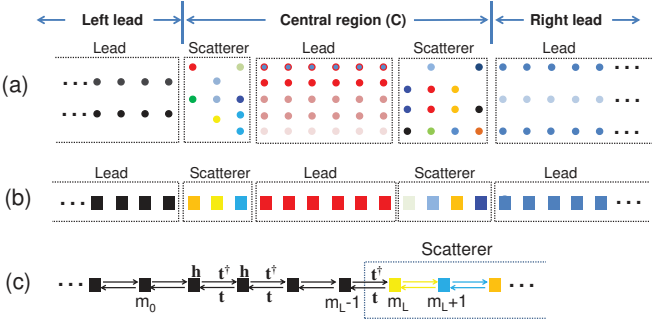}
\caption{(a) A layered
2D structure consisting of multiple periodic slices (i.e., leads) and
disordered slices (i.e., scatterers). (b) Regarding each slice as a unit cell
(filled squares), the structure in (a) becomes a 1D lattice. (c) A
semi-infinite lead connected to a scatterer. The unit cell Hamiltonian (filled
squares) and nearest-neighbor hopping (double arrows) are $m$-independent
inside the lead, but could be disordered inside the scatterer.}%
\label{G_SETUP}%
\end{figure}

We consider a general layered system in the lattice representation. When each
layer is an infinite, periodic repetition of a basic unit, we can make a
Fourier transform to effectively reduce each layer to a single basic unit.
Disorder inside each layer can also be introduced by using a sufficiently
large unit cell and repeating it periodically. Then we can regard each layer
as a finite-size unit cell, so the system becomes a 1D lattice, e.g., by
taking each layer/slice of the structure in Fig. \ref{G_SETUP}(a) as a unit
cell, Fig. \ref{G_SETUP}(a) becomes Fig. \ref{G_SETUP}(b). A general 1D
lattice can always be decomposed into a few nonperiodic regions (referred to
as \textit{scatterers}) consisting of different unit cells sandwiched between
periodic regions (referred to as \textit{leads}) consisting of identical unit
cells; see Fig. \ref{G_SETUP}(b) for an example.

Without losing generality, we consider nearest-neighbor hopping
\footnote{Lattice models with hopping between $p$th nearest neighbors can
always be reduced to nearest-neighbor hopping by retarding $p$ successive unit
cells as one composite unit cell.} and use $M_{m}$ to denote the number of
orthonormal local bases in the $m$th unit cell. In the representation of these
bases, the lattice Hamiltonian is an infinite-dimensional block-tridiagonal
matrix:
\begin{equation}
\mathbf{H}=%
\begin{bmatrix}
\cdots & \cdots & \cdots & \cdots & \cdots & \cdots & \cdots\\
\cdots & \mathbf{H}_{-2,-2} & \mathbf{H}_{-2,-1} & 0 & 0 & 0 & \cdots\\
\cdots & \mathbf{H}_{-2,-1}^{\dagger} & \mathbf{H}_{-1,-1} & \mathbf{H}_{-1,0}
& 0 & 0 & \cdots\\
\cdots & 0 & \mathbf{H}_{-1,0}^{\dagger} & \mathbf{H}_{0,0} & \mathbf{H}_{0,1}
& 0 & \cdots\\
\cdots & 0 & 0 & \mathbf{H}_{0,1}^{\dagger} & \mathbf{H}_{1,1} &
\mathbf{H}_{1,2} & \cdots\\
\cdots & 0 & 0 & 0 & \mathbf{H}_{1,2}^{\dagger} & \mathbf{H}_{2,2} & \cdots\\
\cdots & \cdots & \cdots & \cdots & \cdots & \cdots & \cdots
\end{bmatrix}
, \label{HAMIL}%
\end{equation}
consisting of the $M_{m}\times M_{m+1}$ hopping matrix $\mathbf{H}_{m,m+1}$
between neighboring unit cells, its Hermitian conjugate $\mathbf{H}%
_{m+1,m}=\mathbf{H}_{m,m+1}^{\dagger}$, and the $M_{m}\times M_{m}$
Hamiltonian matrix $\mathbf{H}_{m,m}$ of the $m$th unit cell. In a lead,
$\mathbf{H}_{m,m}=\mathbf{h}$ and $\mathbf{H}_{m,m+1}=\mathbf{t}$ are
independent of $m$. In a scatterer, $\mathbf{H}_{m,m}$ and $\mathbf{H}%
_{m,m+1}$ could dependent on $m$ arbitrarily. Here, as a convention, the
region of the scatterers is chosen such that the hopping between the
lead and the surface of a scatterer is \textit{the same} as that inside this lead,
e.g., $\mathbf{H}_{m_{L}-1,m_{L}}=\mathbf{t}$ in Fig. \ref{G_SETUP}(c), where
$\mathbf{t}$ is the hopping inside the left lead. Except for the Hermiticity
requirement $\mathbf{H}=\mathbf{H}^{\dagger}$, the lattice Hamiltonian is
completely general. The (retarded) GF of the layered system is an
infinite-dimensional matrix: $\mathbf{G}(E)\equiv(z-\mathbf{H})^{-1}$, where
$z\equiv E+i0^{+}$. The GF from the unit cell $m_{0}$ to the unit cell $m$ is
an $M_{m}\times M_{m_{0}}$ matrix and corresponds to the $(m,m_{0})$ block of
$\mathbf{G}(E)$, i.e., $\mathbf{G}_{m,m_{0}}(E)\equiv\lbrack(z-\mathbf{H}%
)^{-1}]_{m,m_{0}}$. Hereafter we consider a fixed energy $E$ or $z\equiv
E+i0^{+}$ and omit this argument for brevity.

To highlight the distinguishing features of our approach and introduce
relevant concepts, we first review the commonly used RGF method before
presenting our idea.

\subsection{Recursive Green's function method}

The idea of the RGF is to build up the entire system out of disconnected subsystems
by the Dyson equation. Let us start from two disconnected subsystems $A$ and
$B$ characterized by the Hamiltonian $\mathbf{H}^{(A)}$ and $\mathbf{H}%
^{(B)}$, respectively. The (retarded) GFs of each subsystem are $\mathbf{G}%
^{(A)}\equiv(z-\mathbf{H}^{(A)})^{-1}$ and $\mathbf{G}^{(B)}\equiv
(z-\mathbf{H}^{(B)})^{-1}$. Next we connect the interface (denoted by $a$) of
$A$ and the interface (denoted by $b$) of $B$ by local couplings
$\mathbf{V}_{ab}$ and $\mathbf{V}_{ba}$. Thus the Dyson equation gives the GF
\begin{equation}
\mathbf{G}=%
\begin{bmatrix}
\mathbf{G}_{AA} & \mathbf{G}_{AB}\\
\mathbf{G}_{BA} & \mathbf{G}_{BB}%
\end{bmatrix}
\end{equation}
of the connected system in terms of the GFs of each subsystem
\cite{VelevJPC2004}:
\begin{subequations}
\label{DSEQ1}%
\begin{align}
\mathbf{G}_{AA}  &  =((\mathbf{G}^{(A)})^{-1}-\mathbf{V}_{ab}\mathbf{G}%
_{bb}^{(B)}\mathbf{V}_{ba})^{-1},\label{DSEQ1_AA}\\
\mathbf{G}_{BB}  &  =((\mathbf{G}^{(B)})^{-1}-\mathbf{V}_{ba}\mathbf{G}%
_{aa}^{(A)}\mathbf{V}_{ab})^{-1},\\
\mathbf{G}_{BA}  &  =\mathbf{G}_{Bb}^{(B)}\mathbf{V}_{ba}\mathbf{G}_{aA},\\
\mathbf{G}_{AB}  &  =\mathbf{G}_{Aa}^{(A)}\mathbf{V}_{ab}\mathbf{G}_{bB},
\end{align}
or vice versa:
\end{subequations}
\begin{subequations}
\label{DSEQ2}%
\begin{align}
\mathbf{G}^{(A)}  &  =\mathbf{G}_{AA}-\mathbf{G}_{Ab}\mathbf{V}_{ba}%
(1+\mathbf{G}_{ab}\mathbf{V}_{ba})^{-1}\mathbf{G}_{aA},\\
\mathbf{G}^{(B)}  &  =\mathbf{G}_{BB}-\mathbf{G}_{Ba}\mathbf{V}_{ab}%
(1+\mathbf{G}_{ba}\mathbf{V}_{ab})^{-1}\mathbf{G}_{bB}.
\end{align}
Equation (\ref{DSEQ1}) is the key to building up the entire system out of
disconnected subsystems, while Eq. (\ref{DSEQ2}) can be used to calculate the
GF of each subsystem when the GF of the connected system is known (e.g., if
the connected system is infinite and periodic \cite{SettnesPRB2015}). The
first two equations of Eq. (\ref{DSEQ1}) show that if we focus on one
subsystem (say $A$), the presence of the other subsystem $B$\ amounts to a
self-energy correction to the interface of $A$:\ $\mathbf{H}_{a,a}%
^{(A)}\rightarrow\mathbf{H}_{a,a}^{(A)}+\mathbf{V}_{ab}\mathbf{G}_{bb}%
^{(B)}\mathbf{V}_{ba}$.

\subsubsection{General procedures of RGF}

In RGF, to calculate the conductance of the general layered system as
described earlier [Eq. (\ref{HAMIL})], the system is first partitioned into the semi-infinite left lead $L$ (unit cells $m\leq0$), the central region $C$ (unit cells $1\leq m\leq N$), and the semi-infinite right lead $R$ (unit cells $m\geq N+1$). The \textit{entire} central region $C$ is regarded  as a scatterer [see Fig.
\ref{G_SETUP}(a) for an example], so the GF is
\end{subequations}
\begin{equation}
\mathbf{G}=%
\begin{bmatrix}
\mathbf{G}_{LL} & \mathbf{G}_{LC} & \mathbf{G}_{LR}\\
\mathbf{G}_{CL} & \mathbf{G}_{CC} & \mathbf{G}_{CR}\\
\mathbf{G}_{RL} & \mathbf{G}_{RC} & \mathbf{G}_{RR}%
\end{bmatrix}
.
\end{equation}
Let us use $\mathbf{H}^{(C)}$ for the Hamiltonian of the central region, and
$\mathbf{H}^{(p)}$ $(p=L,R)$ for the Hamiltonian of the lead $p$, as
characterized by the unit cell Hamiltonian $\mathbf{h}_{p}$ and
nearest-neighbor hopping $\mathbf{t}_{p}=\mathbf{H}_{m,m+1}^{(p)}$. Then the
central region part of the GF is computed from
\begin{equation}
\mathbf{G}_{CC}=(z-\mathbb{H})^{-1}, \label{GCC}%
\end{equation}
where $\mathbb{H}$ is the effective central region Hamiltonian: it equals
$\mathbf{H}^{(C)}$ in the interior of $C$, but differs from $\mathbf{H}^{(C)}$
at the two surface unit cells:
\begin{subequations}
\label{HCC}%
\begin{align}
\mathbb{H}_{1,1}  &  =\mathbf{H}_{1,1}+\mathbf{\Sigma}^{(L)},\\
\mathbb{H}_{N,N}  &  =\mathbf{H}_{N,N}+\mathbf{\Sigma}^{(R)},
\end{align}
due to self-energy corrections from the left and right leads:%
\end{subequations}
\begin{align}
\mathbf{\Sigma}^{(L)}  &  =\mathbf{t}_{L}^{\dagger}\mathbf{G}_{\mathrm{s}%
}^{(L)}\mathbf{t}_{L},\\
\mathbf{\Sigma}^{(R)}  &  =\mathbf{t}_{R}\mathbf{G}_{\mathrm{s}}%
^{(R)}\mathbf{t}_{R}^{\dagger}\mathbf{.}%
\end{align}
Here $\mathbf{G}_{\mathrm{s}}^{(L)}=[(z-\mathbf{H}^{(L)})^{-1}]_{0,0}$ is the
GF of the left lead at the right surface, and $\mathbf{G}_{\mathrm{s}}%
^{(R)}=[(z-\mathbf{H}^{(R)})^{-1}]_{N+1,N+1}$ is the GF of the right lead at
the left surface, so they are referred to as \textit{surface GFs} in the
literature. Finally, to compute the linear conductance, we need to set
$E=E_{F}$ and use the Landauer-Buttiker formula
\cite{LandauerJRD1957,LandauerPM1970,BuettikerPRB1985} $\sigma=(e^{2}%
/h)T(E_{F})$, where \cite{CaroliJPC1971}
\begin{equation}
T(E_{F})=\mathrm{Tr}\mathbf{\Gamma}^{(R)}\mathbf{G}_{N,1}\mathbf{\Gamma}%
^{(L)}(\mathbf{G}_{N,1})^{\dagger} \label{TEF}%
\end{equation}
and $\mathbf{\Gamma}^{(\alpha)}\equiv i(\mathbf{\Sigma}^{(\alpha)}-h.c.)$.
Note that Eq. (\ref{TEF}) only involves $\mathbf{G}_{N,1}$, the $(N,1)$ block
of $\mathbf{G}_{CC}$. Instead of direct matrix inversion [see Eq. (6)], $\mathbf{G}_{N,1}$
can be computed by building up the central region layer by layer
\cite{FerryBook1997} through Eq. (\ref{DSEQ1}). Let us use $\mathbf{G}^{(n)}$
$(n=1,2,\cdots,N$) to denote the GF of the subsystem consisting of the unit
cells $m=1,2,\cdots,n$. The RGF starts from $\mathbf{G}^{(1)}=(z-\mathbb{H}%
_{1,1})^{-1}$, first uses the iteration
\begin{equation}
\mathbf{G}_{n,n}^{(n)}=(z-\mathbb{H}_{n,n}-\mathbf{H}_{n,n-1}\mathbf{G}%
_{n-1,n-1}^{(n-1)}\mathbf{H}_{n-1,n})^{-1} \label{RECURSIVE1}%
\end{equation}
to obtain $\{\mathbf{G}_{n,n}^{(n)}\}$, and then uses the iteration
\begin{equation}
\mathbf{G}_{1,n}^{(n)}=\mathbf{G}_{1,n-1}^{(n-1)}\mathbf{H}_{n-1,n}%
\mathbf{G}_{n,n}^{(n)} \label{RECURSIVE2}%
\end{equation}
to obtain $\mathbf{G}_{1,N}=\mathbf{G}_{1,N}^{(N)}$. The number of iterations
and hence the time cost of the above recursive algorithm scales linearly with
the length of the scattering region.

Equation\ (\ref{TEF}) gives the total transmission probability, i.e., the sum
of the transmission probabilities of all channels. To identify the
contributions from each individual transmission channels, it is necessary to
use the GF to construct the transmission amplitude $S_{\beta,\alpha}^{(RL)}$
from the $\alpha$th traveling channel in the lead $L$ to the $\beta$th
traveling channel in the lead $R$ through the Fisher-Lee relations
\cite{FisherPRB1981,StoneIJRD1988,BarangerPRB1989,SolsAP1992,SanvitoPRB1999,KhomyakovPRB2005,WimmerThesis2009}
and then sum over all the traveling channels:%
\begin{equation}
T(E_{F})=\sum_{\alpha\beta\in\mathrm{traveling}}|S_{\beta,\alpha}^{(RL)}|^{2}.
\label{TEF2}%
\end{equation}
Alternatively, it is also possible to calculate the transmission amplitudes
$\{S_{\beta,\alpha}^{(RL)}\}$ (and more generally the entire scattering
matrix) by directly calculating the scattering of an incident traveling wave
through the wave function mode matching approach
\cite{AndoPRB1991,NikolicPRB1994,XiaPRB2006}. The equivalence between Eqs.
(\ref{TEF}) and (\ref{TEF2}), which establishes a connection between the GF
approach and the wave function mode matching approach, is well known for a
single parabolic band
\cite{FisherPRB1981,StoneIJRD1988,BarangerPRB1989,SolsAP1992,DattaBook1995,FerryBook1997}%
. For a general lattice model, there was suspicion \cite{KrsticPRB2002} that
Eq. (\ref{TEF2}) was incomplete since the GF in Eq. (\ref{TEF}) includes both
traveling waves and evanescent waves, while Eq. (\ref{TEF2}) only includes the
contributions from traveling waves. Later, a rigorous equivalence proof was
provided by Khomyakov \textit{et al.} \cite{KhomyakovPRB2005} and others
\cite{WimmerThesis2009}, but the presence of the evanescent states does
suggest that the GF is not completely equivalent to the unitary scattering matrix.

There are still two remaining issues:\ the calculation of the self-energies
$\mathbf{\Sigma}^{(L,R)}$ (or equivalently the surface GFs $\mathbf{G}%
_{\mathrm{s}}^{(L,R)}$)\ and a proper definition of the scattering channels
and the transmission amplitudes $S_{\beta,\alpha}^{(RL)}$.

\subsubsection{Self-energies: recursive method and eigenmode method}

The numerical algorithms for computing $\mathbf{\Sigma}^{(L,R)}$ or
equivalently the surface GFs $\mathbf{G}_{\mathrm{s}}^{(L,R)}$ can be
classified into two groups: recursive methods and eigenmode methods (see
\noindent Ref. \onlinecite{VelevJPC2004} for a review). The former calculates
an approximate surface GF through some recursive relations, while the latter
provides exact---within the numerical precision---closed-form solutions to
the surface GF.

The idea of the recursive methods is to split the left lead into the surface
unit cell $m=0$ (subsystem $A$) and the remaining part (subsystem $B$); the
Dyson equation [Eq. (\ref{DSEQ1_AA})] gives the recursive relation
\begin{eqnarray}
\mathbf{G}_{\mathrm{s}}^{(L)}&=&(z-\mathbf{h}_{L}-\mathbf{t}_{L}^{\dagger
}\mathbf{G}_{\mathrm{s}}^{(L)}\mathbf{t}_{L})^{-1} \notag \\
\Leftrightarrow
\mathbf{\Sigma}^{(L)}&=&\mathbf{t}_{L}^{\dagger}(z-\mathbf{h}_{L}-\mathbf{\Sigma
}^{(L)})^{-1}\mathbf{t}_{L}.
\end{eqnarray}
Similarly, by splitting the right lead into the surface unit cell $m=N+1$
(subsystem $A$) and the remaining part (subsystem $B$), Eq. (\ref{DSEQ1_AA})
gives the recursive relation%
\begin{eqnarray}
\mathbf{G}_{\mathrm{s}}^{(R)}&=&(z-\mathbf{h}_{R}-\mathbf{t}_{R}\mathbf{G}%
_{\mathrm{s}}^{(R)}\mathbf{t}_{R}^{\mathbf{\dagger}})^{-1} \notag
\\ \Leftrightarrow
\mathbf{\Sigma}^{(R)}&=&\mathbf{t}_{R}(z-\mathbf{h}_{R}-\mathbf{\Sigma}%
^{(R)})^{-1}\mathbf{t}_{R}^{\dagger}.
\end{eqnarray}
Thus the surface GFs and self-energies can be obtained by simple or more
efficient iteration techniques \noindent\cite{SanchoJPF1985,SanchoJPFMP1984}.

The eigenmode method has been derived independently several times
\cite{LeePRB1981,AndoPRB1991,NikolicPRB1994,UmerskiPRB1997,SanvitoPRB1999,KrsticPRB2002,RochaPRB2006}
and has been shown to be superior in accuracy and performance
\cite{UmerskiPRB1997} compared to the recursive methods. The central results
are explicit expressions for the self-energies:
\begin{subequations}
\label{SIGMA}%
\begin{align}
\mathbf{\Sigma}^{(L)}  &  =\mathbf{t}_{L}^{\dagger}(\mathbf{P}_{-}^{(L)}%
)^{-1}\mathbf{,}\\
\mathbf{\Sigma}^{(R)}  &  =\mathbf{t}_{R}\mathbf{P}_{+}^{(R)},
\end{align}
and surface GFs:%
\end{subequations}
\begin{align}
\mathbf{G}_{\mathrm{s}}^{(L)}  &  =(z-\mathbf{h}_{L}-\mathbf{t}_{L}^{\dagger
}(\mathbf{P}_{-}^{(L)})^{-1})^{-1}=(\mathbf{t}_{L}\mathbf{P}_{-}^{(L)}%
)^{-1},\\
\mathbf{G}_{\mathrm{s}}^{(R)}  &  =(z-\mathbf{h}_{R}-\mathbf{t}_{R}%
\mathbf{P}_{+}^{(R)})^{-1}=\mathbf{P}_{+}^{(R)}(\mathbf{t}_{R}^{\dagger}%
)^{-1},
\end{align}
in terms of the (retarded) \textit{propagation matrices} $\mathbf{P}_{\pm
}^{(L,R)}$ (also referred to as Bloch matrices \cite{KhomyakovPRB2005} or
amplitude transfer matrices \cite{VelevJPC2004} in the literature), which can be
constructed from the (retarded) \textit{eigenmodes} of each lead.

Now we introduce the propagation matrices and eigenmodes in some detail, since
they will play a central role in our GF approach. Let us consider a lead
characterized by the unit cell Hamiltonian $\mathbf{h}$ and nearest-neighbor
hopping matrix $\mathbf{t}=\mathbf{H}_{m,m+1}$. The wave propagation in this
lead is governed by the uniform Schr\"{o}dinger equation%
\begin{equation}
-\mathbf{t}^{\dagger}|\Phi(m-1)\rangle+(z-\mathbf{h})|\Phi(m)\rangle
-\mathbf{t}|\Phi(m+1)\rangle=0, \label{SE}%
\end{equation}
where $z\equiv E+i0^{+}$. Imposing the Bloch symmetry $|\Phi(m)\rangle
=e^{ikma}|\Phi\rangle$ ($a$ is the thickness of each unit cell) gives
\begin{equation}
z|\Phi\rangle=(e^{-ika}\mathbf{t}^{\dagger}+\mathbf{h}+e^{ika}\mathbf{t}%
)|\Phi\rangle\equiv\mathbf{H}(k)|\Phi\rangle\label{MEQ1}%
\end{equation}
for the eigenvector $|\Phi\rangle$. For an infinite lead, the wave function
$|\Phi(m)\rangle$ must remain finite at $m\rightarrow\pm\infty$. This natural
boundary condition dictates $k$ to be real, so that Eq. (\ref{MEQ1}) gives $M$
real energy bands of the lead, where $M$ is the number of basis states in each
unit cell of this lead. For certain complex $k$'s, the energies could still be
real, which form the complex energy bands of the lead.

Conversely, given the energy $E$ and without imposing any boundary conditions,
Eq. (\ref{SE}) or (\ref{MEQ1}) can be solved to yield $2M\ $(retarded)
eigenmodes $\{k,|\Phi\rangle\}$ (see Appendix \ref{APPEND_BULKMODE})
\cite{AndoPRB1991,NikolicPRB1994,KhomyakovPRB2005,XiaPRB2006}, where the wave
vector $k$ could be either real (i.e., \textit{traveling} modes)\ or complex
(i.e., \textit{evanescent} modes). The eigenmodes are just the collection of
eigenstates on the energy shell $E$ in the real and complex energy bands of
the lead. As a convention, each eigenvector $|\Phi\rangle$ should be
normalized to unity, but different eigenvectors are not necessarily
orthogonal. For a traveling eigenmodes with wave vector $k$ and eigenvector
$|\Phi\rangle$, its group velocity is
\begin{equation}
v=\partial_{k}\langle\Phi|\mathbf{H}(k)|\Phi\rangle=-2a\operatorname*{Im}%
\langle\Phi|\mathbf{t}e^{ika}|\Phi\rangle, \label{VELOCITY}%
\end{equation}
where $\langle\Phi|$ is the conjugate transpose of $|\Phi\rangle$, i.e., an
$M$-component row vector. Then the $2M$ eigenmodes are classified into $M$
right-going ones and $M$ left-going ones: the former consist of traveling
modes with a positive group velocity and evanescent modes decaying
exponentially to the right (i.e., $\operatorname{Im}k>0$), while the latter
consist of traveling modes with a negative group velocity and evanescent
modes decaying exponentially to the left (i.e., $\operatorname{Im}k<0$). For
clarity, we denote the $M$ right-going eigenmodes as $\{k_{+,\alpha}%
,|\Phi_{+,\alpha}\rangle\}$ and the $M$ left-going eigenmodes as
$\{k_{-,\alpha},|\Phi_{-,\alpha}\rangle\}$, where $\alpha=1,2,\cdots,M$. For every right-going evanescent
mode $(+,\alpha)$ with wave vector $k_{+,\alpha}$, there is always a
left-going evanescent mode $(-,\alpha)$ with wave vector $k_{-,\alpha}$
$=k_{+,\alpha}^{\ast}$ \cite{MolinariJPA1997,SanvitoPRB1999,KhomyakovPRB2005}.

The propagation matrix $\mathbf{P}_{s}$ for left-going $(s=-$) or right-going
($s=+$) waves is constructed as
\cite{AndoPRB1991,NikolicPRB1994,KhomyakovPRB2005,XiaPRB2006}
\begin{equation}
\mathbf{P}_{s}\equiv\mathbf{U}_{s}%
\begin{bmatrix}
e^{ik_{s,1}a} &  & \\
& \ddots & \\
&  & e^{ik_{s,M}a}%
\end{bmatrix}
\mathbf{U}_{s}^{-1}, \label{PM}%
\end{equation}
where $\mathbf{U}_{s}\equiv\lbrack|\Phi_{s,1}\rangle,\cdots,|\Phi_{s,M}%
\rangle]$ (i.e., its $\alpha$th column is $|\Phi_{s,\alpha}\rangle$). The
propagation matrices are standard tools in the wave function mode matching
approach \cite{AndoPRB1991,NikolicPRB1994,KhomyakovPRB2005,XiaPRB2006} to
describe wave propagation; e.g., a general right-going wave that obeys Eq.
(\ref{SE}) can be written as $|\Phi(m)\rangle=\mathbf{P}_{+}^{m}%
|\Phi(0)\rangle$, while a general left-going wave obeying Eq. (\ref{SE}) can
be written as $|\Phi(m)\rangle=\mathbf{P}_{-}^{m}|\Phi(0)\rangle$.

\subsubsection{Scattering channels and Fisher-Lee relations}

Since the GF $\mathbf{G}(E)$ describes the scattering of both traveling and
evanescent eigenmodes by the central region, while the unitary scattering
matrix $\mathbf{S}(E)$ describes the scattering of traveling eigenmodes only,
it is possible to construct $\mathbf{S}(E)$ in terms of $\mathbf{G}(E)$, i.e.,
the Fisher-Lee-type relations. Compared with the Caroli's expression [Eq.
(\ref{TEF})] that gives the total transmission probability, the Fisher-Lee relations further provide information about the
scattering of each individual traveling eigenmode. For a general lattice
model, different eigenmodes $\{|\Phi_{s,\alpha}\rangle\}$ are not orthogonal;
then for each lead, it is necessary to introduce $2M$ (retarded) dual vectors
$\{|\phi_{s,\alpha}\rangle\}$ through
\cite{SanvitoPRB1999,KhomyakovPRB2005,WimmerThesis2009}%
\begin{equation}%
\begin{bmatrix}
\langle\phi_{s,1}|\\
\vdots\\
\langle\phi_{s,M}|
\end{bmatrix}
\equiv\mathbf{U}_{s}^{-1},
\end{equation}
where $M$ is the number of bases in each unit cell of this lead. In general,
different left-going (right-going)\ eigenvectors are not orthogonal, so
$\mathbf{U}_{s}$ is not necessarily unitary and $|\phi_{s,\alpha}\rangle$ is
not necessarily equal to $|\Phi_{s,\alpha}\rangle$, but we always have the
orthonormalization and completeness relations
\begin{align}
\langle\phi_{s,\alpha}|\Phi_{s,\beta}\rangle &  =\langle\Phi_{s,\alpha}%
|\phi_{s,\beta}\rangle=\delta_{\alpha,\beta},\label{ORTHO}\\
\sum_{\alpha}|\Phi_{s,\alpha}\rangle\langle\phi_{s,\alpha}|  &  =\sum_{\alpha
}|\phi_{s,\alpha}\rangle\langle\Phi_{s,\alpha}|=\mathbf{I}, \label{COMPLETE}%
\end{align}
which follow from $\mathbf{U}_{s}\mathbf{U}_{s}^{-1}=\mathbf{U}_{s}%
^{-1}\mathbf{U}_{s}=\mathbf{I}$ ($\mathbf{I}$ is the identity matrix). By
inserting the completeness relation, any column vector $|\Phi\rangle$ can be
expanded as a linear combination of either the $M$ left-going eigenvectors or
the $M$ right-going\ eigenvectors as $|\Phi\rangle=\sum_{\alpha}c_{s,\alpha
}|\Phi_{s,\alpha}\rangle$ with $c_{s,\alpha}=\langle\phi_{s,\alpha}%
|\Phi\rangle$. In terms of the eigenmodes and their dual vectors, Eq.
(\ref{PM}) can be written as%
\begin{equation}
\mathbf{P}_{s}\equiv\sum_{\alpha}e^{ik_{s,\alpha}a}|\Phi_{s,\alpha}%
\rangle\langle\phi_{s,\alpha}|, \label{P_EXPAND}%
\end{equation}
which has a clear physical interpretation. For example, a general right-going
wave is%
\begin{equation}
|\Phi(m)\rangle=\mathbf{P}_{+}^{m}|\Phi(0)\rangle=\sum_{\alpha}e^{ik_{+,\alpha
}ma}|\Phi_{+,\alpha}\rangle\langle\phi_{+,\alpha}|\Phi(0)\rangle,
\end{equation}
i.e., $|\Phi(0)\rangle$ is first expanded as a linear combination of
right-going eigenmodes $|\Phi(0)\rangle=\sum_{\alpha}|\Phi_{+,\alpha}%
\rangle\langle\phi_{+,\alpha}|\Phi(0)\rangle$ and then each right-going
eigenmode propagates as $|\Phi_{+,\alpha}\rangle\rightarrow e^{ik_{+,\alpha
}ma}|\Phi_{+,\alpha}\rangle$.

Now the scattering channel can be labeled by the eigenmodes, which could be
either traveling or evanescent. The scattering matrix $\mathbf{S}(E)$ provides
a complete description for the scattering from one traveling eigenmode into
another traveling eigenmode. For a general lattice model, the Fisher-Lee
relations allow us to construct the scattering matrix from the lattice GF,
e.g., the transmission amplitude from the right-going eigenmode $|\Phi
_{+,\alpha}^{(L)}\rangle$ of the left lead into the right-going eigenmode
$|\Phi_{+,\beta}^{(R)}\rangle$ of the right lead is
\cite{AndoPRB1991,SanvitoPRB1999,KhomyakovPRB2005,WimmerThesis2009}:%
\begin{equation}
S_{\beta,\alpha}^{(R,L)}|_{\alpha,\beta\in\mathrm{traveling}}=\sqrt
{\frac{v_{+,\beta}^{(R)}/a_{R}}{v_{+,\alpha}^{(L)}/a_{L}}}\langle\phi
_{+,\beta}^{(R)}|\mathbf{G}_{N,1}(\mathbf{g}^{(L)})^{-1}|\Phi_{+,\alpha}%
^{(L)}\rangle, \label{SUNITARY_RL}%
\end{equation}
while the transmission amplitude from the left-going eigenmode $|\Phi
_{-,\alpha}^{(R)}\rangle$ of the right lead into the left-going eigenmode
$|\Phi_{-,\beta}^{(L)}\rangle$ of the left lead is
\cite{AndoPRB1991,SanvitoPRB1999,KhomyakovPRB2005,WimmerThesis2009}%
\begin{equation}
S_{\beta,\alpha}^{(L,R)}|_{\alpha,\beta\in\mathrm{traveling}}=\sqrt
{\frac{v_{-,\beta}^{(L)}/a_{L}}{v_{-,\alpha}^{(R)}/a_{R}}}\langle\phi
_{-,\beta}^{(L)}|\mathbf{G}_{1,N}(\mathbf{g}^{(R)})^{-1}|\Phi_{-,\alpha}%
^{(R)}\rangle, \label{SUNITARY_LR}%
\end{equation}
where $a_{p}$ and $\mathbf{g}^{(p)}$ are, respectively, the unit cell
thickness and \textit{free GF} of the lead $p$ [see Eq. (\ref{G00})], and
$v_{s,\alpha}^{(p)}$ is the group velocity [see Eq. (\ref{VELOCITY})] of the
traveling eigenmode $|\Phi_{s,\alpha}^{(p)}\rangle$ in the lead $p$.

\subsection{Our GF approach: Key ideas}

Let us assume that there is a local excitation at the unit cell $m_{0}$, as
described by an $M_{m_{0}}$-dimensional column vector $|\Phi_{\mathrm{loc}%
}\rangle_{m_{0}}$. This excitation generates a casual scattering wave
$|\Phi(m)\rangle$ that has an energy $E$ and obeys the Schr\"{o}dinger
equation with a local source at $m_{0}$:%
\begin{align}
  -\mathbf{H}_{m,m-1}|\Phi(m-1)\rangle+(z-\mathbf{H}_{m,m})|\Phi
(m)\rangle\nonumber\\
  -\mathbf{H}_{m,m+1}|\Phi(m+1)\rangle=\delta_{m,m_{0}}|\Phi_{\mathrm{loc}}\rangle_{m_{0}}.\label{EOM}%
\end{align}
The solution is given by
\begin{equation}
|\Phi(m)\rangle=\mathbf{G}_{m,m_{0}}|\Phi_{\mathrm{loc}}\rangle_{m_{0}%
},\label{PHI_GF}%
\end{equation}
e.g., for a unit excitation of the $\alpha$th basis state, as described by
$|\Phi_{\mathrm{loc}}\rangle_{m_{0}}=[0,\cdots,1,0,\cdots,0]^{T}$ (only the
$\alpha$th element is nonzero), Eq. (\ref{PHI_GF}) gives $|\Phi(m)\rangle$ as
the $\alpha$th column of $\mathbf{G}_{m,m_{0}}$.

\begin{figure}[t]
\includegraphics[width=\columnwidth,clip]{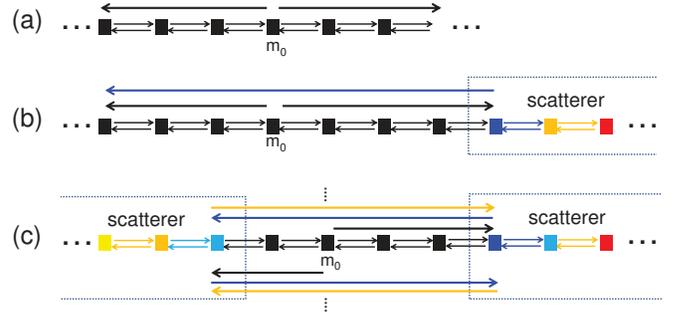}
\caption{Scattering state emanating from a local excitation at $m_{0}$ in an
infinite lead (a), a semi-infinite lead connected to a scatterer on its right
(b), and a finite lead sandwiched between two scatterers. The zeroth-order,
first-order, and second-order partial waves are denoted by black, blue, and
orange arrows, respectively.}%
\label{G_BASICPROCESS}%
\end{figure}

Equation (\ref{PHI_GF}) shows that the GF can be immediately obtained once the
scattering state is determined, e.g., based on physical considerations on how
the local excitation evolves to the scattering state. For example, if the
local excitation $|\Phi_{\mathrm{loc}}\rangle_{m_{0}}$ occurs inside a lead,
then it first generates an outgoing zeroth-order wave consisting of a
left-going one at $m\leq m_{0}$ and a right-going one at $m\geq m_{0}$. For an
infinite lead, there are no scatterers, so this outgoing wave is the total
scattering state [Fig. \ref{G_BASICPROCESS}(a)]. For a semi-infinite lead
connected to a scatterer on its right [Fig. \ref{G_BASICPROCESS}(b)], the
zeroth-order right-going wave will produce a first-order reflection wave, so
the total scattering state in the lead is the sum of the zeroth-order and
first-order waves. More generally, for a finite lead sandwiched between two
scatterers [Fig. \ref{G_BASICPROCESS}(c)], the right-going (left-going)
zeroth-order wave will propagate to the right (left) scatterer and produce a
first-order reflection wave, which in turn will propagate to the left (right)
scatterer and then produce high-order reflection waves. The total scattering
state would be the sum of all these waves.

In contrast to the commonly used RGF that treats the \textit{entire} central region (regarded as a large scatterer) numerically [see Fig. \ref{G_SETUP}(a) for an example], our method need only regard each \textit{truly disordered} region as a scatterer for numerical treatment, while all the periodic subregions [such as the middle lead in Fig. \ref{G_SETUP}(a)] inside the central region can be treated semianalytically by fully utilizing the translational invariance of these subregions. Physically, the wave propagation inside these periodic subregions leads to complicated multiple reflection between different scatterers, which is difficult to handle analytically in the standard RGF technique in which the GF is treated as a matrix. By contrast, in our approach, it is straightforward to perform analytically an infinite summation over all the multiple reflection waves, so that the time cost can be significantly reduced. This approach also provides a
physically transparent expansion of the GF $G(E)$ in terms of a generalized
scattering matrix $\mathcal{S}(E)$, which can be regarded as a reverse of the
well-known Fisher-Lee relations
\cite{FisherPRB1981,StoneIJRD1988,BarangerPRB1989,SolsAP1992,SanvitoPRB1999,KhomyakovPRB2005,WimmerThesis2009} (see Fig. 1).
In the next section, we will establish the procedures for calculating the GF within this framework in a physically transparent way.
\section{Our Green's function approach}

Our GF approach essentially consists of two steps: generation of the
zeroth-order outgoing partial wave by the local excitation and its propagation
in the leads and scattering by the scatterers. The wave function mode matching
approach \cite{AndoPRB1991,NikolicPRB1994,KhomyakovPRB2005,XiaPRB2006} has
developed useful tools to describe the latter process. Below we begin with an
infinite lead, then we consider an infinite system containing a single
scatterer. Finally, we give the analytical construction rule for the GF of a
general layered system containing an arbitrary number of scatterers.

\subsection{Infinite lead}

Suppose that the lead is characterized by the unit cell Hamiltonian
$\mathbf{h}$ and hopping $\mathbf{t}$. The scattering state $|\Phi(m)\rangle$
emanating from a local excitation $|\Phi_{\mathrm{loc}}\rangle_{m_{0}}$ is
determined by the Schr\"{o}dinger equation with a local source at $m_{0}$:
\begin{align}
-\mathbf{t}^{\dagger}|\Phi(m-1)\rangle+(z-\mathbf{h})|\Phi(m)\rangle
-\mathbf{t}|\Phi(m+1)\rangle \notag
\\=\delta_{m,m_{0}}|\Phi_{\mathrm{loc}}%
\rangle_{m_{0}}. \label{EOM_0}%
\end{align}
In either the left region $(m\leq m_{0}-1$) or the right region ($m\geq
m_{0}+1$), the local source vanishes, so the general solution would be a
linear combination of left-going and right-going eigenmodes with energy $E$.
However, by causality considerations (due to the infinitesimal imaginary part
of the energy $z=E+i0^{+}$), the solution in the left (right) region should be
a left-going (right-going) wave [see Fig. \ref{G_BASICPROCESS}(a)]:
\begin{align}
|\Phi(m)\rangle|_{m\leq m_{0}-1}  &  =(\mathbf{P}_{-})^{m-m_{0}}|\Phi
(m_{0})\rangle,\\
|\Phi(m)\rangle|_{m\geq m_{0}+1}  &  =(\mathbf{P}_{+})^{m-m_{0}}|\Phi
(m_{0})\rangle.
\end{align}
Substituting into Eq. (\ref{EOM_0}) gives $|\Phi(m_0)\rangle=\mathbf{g}%
|\Phi_{\mathrm{loc}}\rangle_{m_{0}}$, where
\begin{subequations}
\label{G00}%
\begin{align}
\mathbf{g}  &  \equiv(z-\mathbf{h-t}^{\dagger}\mathbf{P}_{-}^{-1}%
-\mathbf{tP}_{+})^{-1}\label{G00A}\\
&  =[\mathbf{t(P}_{-}-\mathbf{P}_{+})]^{-1}=[\mathbf{t}^{\dagger}%
(\mathbf{P}_{+}^{-1}-\mathbf{P}_{-}^{-1})]^{-1}. \label{G00B}%
\end{align}
Here we have used the equality \cite{KhomyakovPRB2005}%
\end{subequations}
\begin{equation}
E-\mathbf{h=t}^{\dagger}\mathbf{P}_{\pm}^{-1}+\mathbf{tP}_{\pm}
\label{EQUALITY}%
\end{equation}
in arriving at Eq. (\ref{G00B}). From the scattering wave function, we
immediately identify the GF of an infinite lead as
\begin{equation}
\mathbf{g}_{m,m_{0}}\equiv\left\{
\begin{array}
[c]{ll}%
\mathbf{P}_{+}^{m-m_{0}}\mathbf{g\ \ } & (m\geq m_{0}),\\
\mathbf{P}_{-}^{m-m_{0}}\mathbf{g} & (m\leq m_{0}).
\end{array}
\right.  \label{G_INFINITE}%
\end{equation}
This
recovers the previous result \cite{SanvitoPRB1999,KhomyakovPRB2005} obtained
by directly solving the equations of motion of the GF. For convenience,
hereafter we call $\mathbf{g}_{m,m_{0}}$ the \textit{free GF} of the lead
since it describes the generation of the zeroth-order outgoing wave
$|\Phi(m)\rangle=\mathbf{g}_{m,m_{0}}|\Phi_{\mathrm{loc}}\rangle_{m_{0}}$ from
a local excitation inside this lead.

\subsection{Single scatterer}

\begin{figure}[t]
\includegraphics[width=\columnwidth,clip]{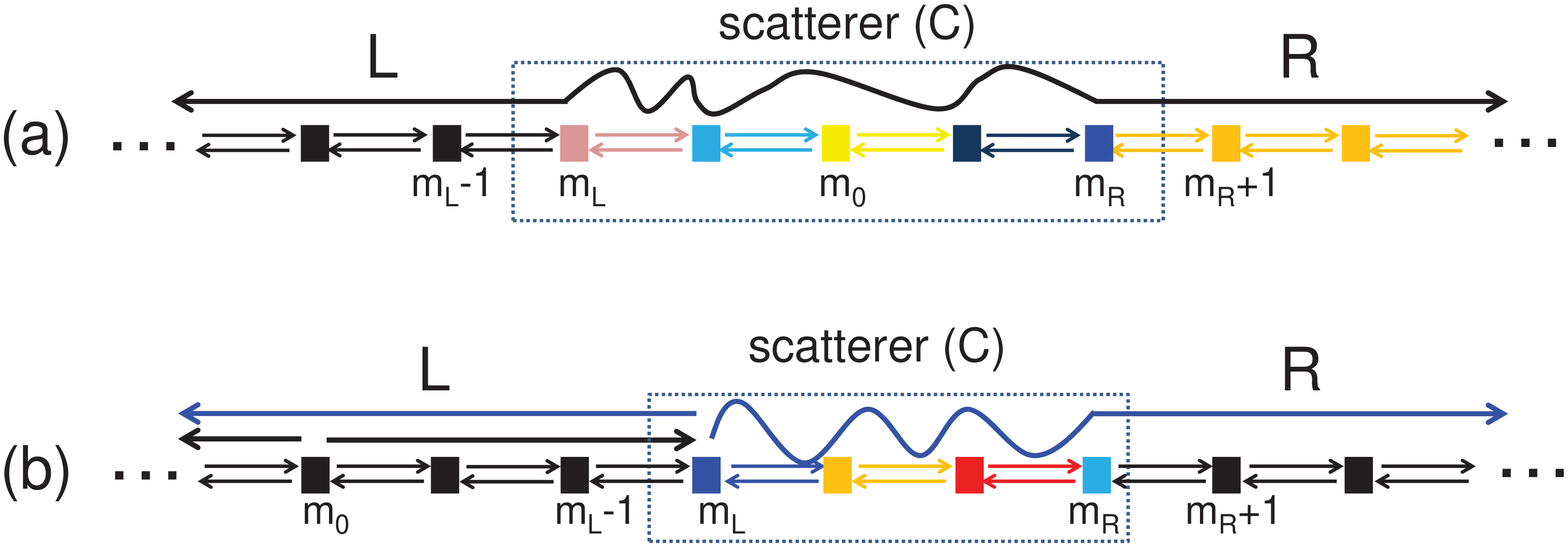}
\caption{Scattering
state emanating from a local excitation inside a scatterer (a) or a lead (b).
The black arrows denote the zeroth-order partial wave and the blue arrows
denote the first-order partial wave due to scattering.}%
\label{G_SINGLE}%
\end{figure}

Let us consider a scatterer $C$ connected to two semi-infinite leads $L$ and
$R$ [Fig. \ref{G_SINGLE}(a)]. The left (right) surface of the scatterer is
$m_{L}$ ($m_{R}$). The unit cell Hamiltonian and hopping inside the left
(right) lead are $\mathbf{h}_{L}$ and $\mathbf{t}_{L}$ ($\mathbf{h}_{R}$ and
$\mathbf{t}_{R}$).

\subsubsection{Local excitation inside the scatterer}

The zeroth-order outgoing wave $|\Phi(m)\rangle$ emanating from a local
excitation $|\Phi_{\mathrm{loc}}\rangle_{m_{0}}$ at $m_{0}\in C$ obeys Eq.
(\ref{EOM}) with $m_{L}\leq m\leq m_{R}$, i.e., inside the scatterer. Inside
the left lead, $|\Phi(m)\rangle$ obeys
\begin{align}
-\mathbf{t}_{L}^{\dagger}|\Phi(m-1)\rangle+(z-\mathbf{h}_{L})|\Phi
(m)\rangle-\mathbf{t}_{L}|\Phi(m+1)\rangle=0
\end{align}
with $m\leq m_{L}-1$. Inside the right lead, $|\Phi(m)\rangle$ obeys%
\begin{equation}
-\mathbf{t}_{R}^{\dagger}|\Phi(m-1)\rangle+(z-\mathbf{h}_{R})|\Phi
(m)\rangle-\mathbf{t}_{R}|\Phi(m+1)\rangle=0
\end{equation}
with $m\geq m_{R}+1$. By causality, the solution in the left (right) lead is a
left-going (right-going) wave [see Fig. \ref{G_SINGLE}(a)]:
\begin{align}
|\Phi(m)\rangle|_{m\in L}  &  =(\mathbf{P}_{-}^{(L)})^{m-m_{L}}|\Phi
(m_{L})\rangle,\\
|\Phi(m)\rangle|_{m\in R}  &  =(\mathbf{P}_{+}^{(R)})^{m-m_{R}}|\Phi
(m_{R})\rangle,
\end{align}
where $\mathbf{P}_{\pm}^{(p)}$ are propagation matrices of lead $p$ [see Eq. (22)].
Substituting $|\Phi(m_{L}-1)\rangle=(\mathbf{P}_{-}^{(L)})^{-1}|\Phi
(m_{L})\rangle$ and $|\Phi(m_{R}+1)\rangle=\mathbf{P}_{+}^{(R)}|\Phi
(m_{R})\rangle$ into Eq. (\ref{EOM}) gives a closed set of equations for
$|\Phi(m)\rangle$ inside the scatterer. The solution is%
\begin{equation}
|\Phi(m)\rangle|_{m\in C}=\mathbb{G}_{m,m_{0}}|\Phi_{\mathrm{loc}}%
\rangle_{m_{0}}, \label{PHI0_SCATTERER}%
\end{equation}
where
\begin{equation}
\mathbb{G}\equiv(z-\mathbb{H})^{-1} \label{GG}%
\end{equation}
and $\mathbb{H}$ is the effective Hamiltonian for the scatterer:\ it is equal
to the scatterer part of the system Hamiltonian $\mathbf{H}$, except for the
two surface unit cells:
\begin{subequations}
\label{HH}
\begin{align}
\mathbb{H}_{m_{L},m_{L}} &  =\mathbf{H}_{m_{L},m_{L}}+\mathbf{t}_{L}^{\dagger
}(\mathbf{P}_{-}^{(L)})^{-1},\\
\mathbb{H}_{m_{R},m_{R}} &  =\mathbf{H}_{m_{R},m_{R}}+\mathbf{t}_{R}%
\mathbf{P}_{+}^{(R)}.
\end{align}
\end{subequations}
Since $\mathbb{G}$ converts a local excitation inside the scatterer into a
scattering state inside the scatterer, we call it the \textit{conversion
matrix} of the scatterer. Comparing Eqs. (\ref{GG}) and (\ref{HH}) to Eqs.
(\ref{GCC}), (\ref{HCC}), and (\ref{SIGMA}), we see that $\mathbb{G}$ is just
the scatterer part of the GF. Actually, from the scattering wave function, we
immediately identify the GF:
\begin{subequations}
\label{G_ALL_S}%
\begin{align}
\mathbf{G}_{m\in C,m_{0}\in C}  &  =\mathbb{G}_{m,m_{0}},\label{GSS}\\
\mathbf{G}_{m\in L,m_{0}\in S}  &  =(\mathbf{P}_{-}^{(L)})^{m-m_{L}}%
\mathbb{G}_{m_{L},m_{0}},\label{GLS}\\
\mathbf{G}_{m\in R,m_{0}\in S}  &  =(\mathbf{P}_{+}^{(R)})^{m-m_{R}}%
\mathbb{G}_{m_{R},m_{0}}. \label{GRS}%
\end{align}
Equation (\ref{GLS}) shows that the local excitation first evolves to an
outgoing wave $\mathbb{G}_{m_{L},m_{0}}|\Phi_{\mathrm{loc}}\rangle_{m_{0}}$ at
the left surface of the scatterer, and then propagates to the unit cell $m$ as
$(\mathbf{P}_{-}^{(L)})^{m-m_{L}}\mathbb{G}_{m_{L},m_{0}}|\Phi_{\mathrm{loc}%
}\rangle_{m_{0}}$. Equation (\ref{GRS}) has a similar physical interpretation.

\subsubsection{Local excitation in the lead}

As shown in Fig. \ref{G_SINGLE}(b), for $m_{0}\in L$, the local excitation
first generates a zeroth-order outgoing wave in the left lead:$\ |\Phi
^{(0)}(m)\rangle=\mathbf{g}_{m,m_{0}}^{(L)}|\Phi_{\mathrm{loc}}\rangle_{m_{0}%
}$, where $\mathbf{g}_{m,m_{0}}^{(p)}$ is the free GF of the lead $p$ [Eq.
(\ref{G_INFINITE})]. Next the right-going partial wave reaches the left
surface of $C$ and evolves into a scattering state $|\Psi(m)\rangle$. For
$m_{0}\in R$, the local excitation first generates a zeroth-order outgoing
wave in the right lead: $|\Phi^{(0)}(m)\rangle=\mathbf{g}_{m,m_{0}}^{(R)}%
|\Phi_{\mathrm{loc}}\rangle_{m_{0}}$. Next the left-going partial wave reaches
the right surface of $C$ and evolves to a scattering state $|\Psi(m)\rangle$.
For either case, the total scattering state $|\Phi(m)\rangle$ emanating from
the local excitation is the sum of the unscattered zeroth-order partial wave
and the scattering state $|\Psi(m)\rangle$. The central issue is to determine
the scattering state emanating from a known incident wave, in a way similar to
the wave function mode matching approach to mesoscopic quantum transport
\cite{AndoPRB1991,NikolicPRB1994,KhomyakovPRB2005,XiaPRB2006}.

First we consider the scattering state $|\Psi(m)\rangle$ emanating from a
right-going incident wave $|\Phi_{\mathrm{in}}(m)\rangle$ in the left lead.
The key observation is that for arbitrary $m_{1}\leq m_{L}$, the local
excitation $|\Phi_{\mathrm{loc}}\rangle_{m_{1}}\equiv(\mathbf{g}^{(L)}%
)^{-1}|\Phi_{\mathrm{in}}(m_{1})\rangle$ generates a right-going partial wave
$|\tilde{\Phi}(m)\rangle|_{m\geq m_{1}}=(\mathbf{P}_{+}^{(L)})^{m-m_{1}}%
|\Phi_{\mathrm{in}}(m_{1})\rangle$ that is equal to $|\Phi_{\mathrm{in}%
}(m)\rangle|_{m\geq m_{1}}$. Therefore, in the region $m\geq m_{1}$, the
scattering state emanating from $|\Phi_{\mathrm{in}}(m)\rangle$ is the same as
the scattering state emanating from this local excitation (see Appendix
\ref{APPEND_EQUIVALENCE} for a rigorous proof). Taking $m_{1}=m_{L}$
immediately gives
\end{subequations}
\begin{equation}
|\Psi(m)\rangle|_{m\in C}=\mathbb{G}_{m,m_{L}}(\mathbf{g}^{(L)})^{-1}%
|\Phi_{\mathrm{in}}(m_{L})\rangle, \label{PHI_S1}%
\end{equation}
i.e., first the incident wave amplitude $|\Phi_{\mathrm{in}}(m_{L})\rangle$ is
converted back to a local excitation $|\Phi_{\mathrm{loc}}\rangle_{m_{L}%
}\equiv(\mathbf{g}^{(L)})^{-1}|\Phi_{\mathrm{in}}(m_{L})\rangle$, then the
conversion matrix $\mathbb{G}$ of the scatterer further converts it to the
total scattering state $|\Psi(m)\rangle|_{m\in C}$ according to Eq.
(\ref{PHI0_SCATTERER}). Inside the left lead, $|\Psi(m)\rangle$ is the sum of
the right-going incident wave and a left-going reflection partial wave
$|\Phi_{\mathrm{r}}(m)\rangle|_{m\in L}=(\mathbf{P}_{-}^{(L)})^{m-m_{L}}%
|\Phi_{\mathrm{r}}(m_{L})\rangle$, where
\begin{align}
|\Phi_{\mathrm{r}}(m_{L})\rangle &  =|\Psi(m_{L})\rangle-|\Phi_{\mathrm{in}%
}(m_{L})\rangle\nonumber\\
&  =[\mathbb{G}_{m_{L},m_{L}}(\mathbf{g}^{(L)})^{-1}-\mathbf{I}]|\Phi
_{\mathrm{in}}(m_{L})\rangle. \label{PHIR_ML}%
\end{align}
Inside the right lead, $|\Psi(m)\rangle$ is the right-going transmission wave:
$|\Psi(m)\rangle|_{m\in R}=(\mathbf{P}_{+}^{(R)})^{m-m_{R}}|\Psi(m_{R}%
)\rangle$, where%
\begin{equation}
|\Psi(m_{R})\rangle=\mathbb{G}_{m_{R},m_{L}}(\mathbf{g}^{(L)})^{-1}%
|\Phi_{\mathrm{in}}(m_{L})\rangle. \label{PHI_MR}%
\end{equation}

Similarly, we can derive the scattering state $|\Psi(m)\rangle$ emanating from
a left-going incident wave $|\Phi_{\mathrm{in}}(m)\rangle$ in the right lead.
Inside the scatterer, the scattering state is
\begin{equation}
|\Psi(m)\rangle|_{m\in S}=\mathbb{G}_{m,m_{R}}(\mathbf{g}^{(R)})^{-1}%
|\Phi_{\mathrm{in}}(m_{R})\rangle, \label{PHI_S2}%
\end{equation}
as if it emanated from a local excitation $|\Phi_{\mathrm{loc}}\rangle_{m_{R}%
}\equiv(\mathbf{g}^{(R})^{-1}|\Phi_{\mathrm{in}}(m_{R})\rangle$ at the right
surface of the scatterer [cf. Eq. (\ref{PHI0_SCATTERER})]. Inside the right
lead, $|\Psi(m)\rangle$ is the sum of the left-going incident wave and a
right-going reflection partial wave $|\Phi_{\mathrm{r}}(m)\rangle|_{m\in
R}=(\mathbf{P}_{+}^{(R)})^{m-m_{R}}|\Phi_{\mathrm{r}}(m_{R})\rangle$, where%
\begin{align}
|\Phi_{\mathrm{r}}(m_{R})\rangle &  =|\Psi(m_{R})\rangle-|\Phi_{\mathrm{in}%
}(m_{R})\rangle\nonumber\\
&  =[\mathbb{G}_{m_{R},m_{R}}(\mathbf{g}^{(R)})^{-1}-\mathbf{I}]|\Phi
_{\mathrm{in}}(m_{R})\rangle. \label{PHIR_MR}%
\end{align}
Inside the left lead, $|\Psi(m)\rangle$ is the left-going transmission
wave:\ $|\Psi(m)\rangle=(\mathbf{P}_{-}^{(L)})^{m-m_{L}}|\Psi(m_{L})\rangle$,
where%
\begin{equation}
|\Psi(m_{L})\rangle=\mathbb{G}_{m_{L},m_{R}}(\mathbf{g}^{(R)})^{-1}%
|\Phi_{\mathrm{in}}(m_{R})\rangle. \label{PHI_ML}%
\end{equation}

Using the above results, the scattering state emanating from the zeroth-order
right-going partial wave in the left lead is given by Eqs. (\ref{PHI_S1}%
)-(\ref{PHI_MR}) with $|\Phi_{\mathrm{in}}(m_{L})\rangle\equiv|\Phi
^{(0)}(m_{L})\rangle$. This allows us to identify the GF
\begin{subequations}
\label{G_ALL_L}%
\begin{align}
\mathbf{G}_{m\in C,m_{0}\in L}  &  =\mathbb{G}_{m,m_{L}}(\mathbf{g}%
^{(L)})^{-1}\mathbf{g}_{m_{L},m_{0}}^{(L)},\label{GSL}\\
\mathbf{G}_{m\in R,m_{0}\in L}  &  =(\mathbf{P}_{+}^{(R)})^{m-m_{R}}%
\mathbb{G}_{m_{R},m_{L}}(\mathbf{g}^{(L)})^{-1}\mathbf{g}_{m_{L},m_{0}}%
^{(L)},\label{GRL}\\
\mathbf{G}_{m\in L,m_{0}\in L}  &  =\mathbf{g}_{m,m_{0}}^{(L)}+(\mathbf{P}%
_{-}^{(L)})^{m-m_{L}}[\mathbb{G}_{m_{L},m_{L}}(\mathbf{g}^{(L)})^{-1}%
-\mathbf{I}]\mathbf{g}_{m_{L},m_{0}}^{(L)}. \label{GLL}%
\end{align}
These expressions have clear physical interpretations; e.g., Eq. (\ref{GSL})
shows that the local excitation $|\Phi_{\mathrm{loc}}\rangle_{m_{0}\in L}$
first evolves to a right-going partial wave and propagates rightward to the
left surface of the scatterer as $\mathbf{g}_{m_{L},m_{0}}^{(L)}%
|\Phi_{\mathrm{loc}}\rangle_{m_{0}\in L}$. There it is converted back to a
local excitation by $(\mathbf{g}^{(L)})^{-1}$, and finally the conversion
matrix of the scatterer $\mathbb{G}_{m,m_{L}}$ further converts it to the
scattering state inside the scatterer. As another example, Eq. (\ref{GLL})
shows that the total scattering wave inside the left lead is the sum of the
zeroth-order partial wave $\mathbf{g}_{m,m_{0}}^{(L)}|\Phi_{\mathrm{loc}%
}\rangle_{m_{0}}$ and the reflection partial wave: first the local excitation
$|\Phi_{\mathrm{loc}}\rangle_{m_{0}\in L}$ evolves to a zeroth-order partial
wave and then propagates rightwards to the left surface of the scatterer as
$\mathbf{g}_{m_{L},m_{0}}^{(L)}|\Phi_{\mathrm{loc}}\rangle_{m_{0}\in L}$, then
$\mathbb{G}_{m_{L},m_{L}}(\mathbf{g}^{(L)})^{-1}-\mathbf{I}$ converts it to
the reflection wave. Finally, $(\mathbf{P}_{-}^{(L)})^{m-m_{L}}$ propagates
this reflection wave leftward to $m$.

Similarly, the scattering state emanating from the zeroth-order left-going
partial wave in the right lead is given by Eqs. (\ref{PHI_S2})-(\ref{PHI_ML})
with $|\Phi_{\mathrm{in}}(m_{R})\rangle\equiv|\Phi^{(0)}(m_{R})\rangle$. This
allows us to identify the GF
\end{subequations}
\begin{subequations}
\label{G_ALL_R}%
\begin{align}
\mathbf{G}_{m\in C,m_{0}\in R}  &  =\mathbb{G}_{m,m_{R}}(\mathbf{g}%
^{(R)})^{-1}\mathbf{g}_{m_{R},m_{0}}^{(R)},\label{GSR}\\
\mathbf{G}_{m\in L,m_{0}\in R}  &  =(\mathbf{P}_{-}^{(L)})^{m-m_{L}}%
\mathbb{G}_{m_{L},m_{R}}(\mathbf{g}^{(R)})^{-1}\mathbf{g}_{m_{R},m_{0}}%
^{(R)},\label{GLR}\\
\mathbf{G}_{m\in R,m_{0}\in R}  &  =\mathbf{g}_{m,m_{0}}^{(R)}+(\mathbf{P}%
_{+}^{(R)})^{m-m_{R}}[\mathbb{G}_{m_{R},m_{R}}(\mathbf{g}^{(R)})^{-1}%
-\mathbf{I}]\mathbf{g}_{m_{R},m_{0}}^{(R)}. \label{GRR}%
\end{align}
These can be interpreted in a similar way to Eqs. (\ref{G_ALL_S}) and
(\ref{G_ALL_L}).

The above results cover previous results as special cases. For example, by
directly solving the equation of motion, Sanvito \textit{et al. }%
\cite{SanvitoPRB1999} and Krsti\'{c} \textit{et al.} \cite{KrsticPRB2002}
obtain the GF of an infinite lead [Eq. (\ref{G_INFINITE})] and a semi-infinite
lead consisting of the unit cells $m\leq0$ ($m\geq0$) [Eq. (\ref{GL_MM0})],
which can be regarded as a single-unit-cell scatterer at $m=0$ connected to a
semi-infinite left (right) lead. Khomyakov \textit{et al. }%
\cite{KhomyakovPRB2005} further obtained the GF across a single scatterer [Eq.
(\ref{GRL})]. A sharp interface between a semi-infinite left lead and a
semi-infinite right lead can also be regarded as a single-unit-cell scatterer
connected to two semi-infinite leads. For reference, the explicit expressions
of the GFs for these simple cases are given in Appendix \ref{APPEND_EXAMPLE}.

\subsection{Multiple scatterers}

A general layered system containing an arbitrary number of scatterers can be
regarded as a \textit{composite} scatterer connected to one or two
semi-infinite leads, e.g., a scatterer $B$ connected to a finite lead $L$ and
a semi-infinite lead $R$ can be regarded as a composite scatterer $C=(A+B)$
connected to one semi-infinite right lead [Fig. \ref{G_COMPOSITE_SCATTERER}%
(a)], while two scatterers sandwiched between three leads can be regarded a
composite scatterer $C=(A+B)$ connected to two semi-infinite leads [Fig.
\ref{G_COMPOSITE_SCATTERER}(b)]. Therefore, we can use Eqs. (\ref{G_ALL_S}),
(\ref{G_ALL_L}), and (\ref{G_ALL_R}) to obtain the GF of the entire system
once the conversion matrix of the composite scatterer is known. In the RGF
method, the conversion matrix (which coincides with the GF of the infinite system within the composite scatterer) is calculated by a numerical iteration algorithm that builds up the
composite scatterer slice by slice, thus the time cost
increases linearly with the total length of the composite scatterer. Here the
physical transparency of our approach allows us to treat the multiple
reflection between different scatterers analytically, so that the conversion
matrix of a composite scatterer can be obtained by combining the
conversion matrices of the constituent scatterers analytically with a significantly reduced
time cost. The basic step is the combination of the conversion matrices
$\mathbb{G}^{(A)}$ and $\mathbb{G}^{(B)}$ of two scatterers $A$ and $B$ into
the conversion matrix $\mathbb{G}$ of a composite scatterer $C\equiv(A+B)$.

\subsubsection{Combining conversion matrices}

\begin{figure}[t]
\includegraphics[width=\columnwidth,clip]{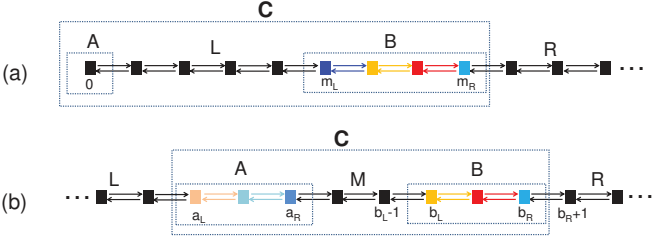}
\caption{(a)
A scatterer $B$ connected to a finite left lead $L$ and a semi-infinite right
lead $R$ can be regarded as a composite scatterer $C$ connected to one
semi-infinite right lead $R$. (b) Two scatterers sandwiched between three
leads $L,M,R$ can be regarded as a composite scatterer $C$ connected to two
semi-infinite leads $L$ and $R$.}%
\label{G_COMPOSITE_SCATTERER}%
\end{figure}

As shown in Fig. \ref{G_COMPOSITE_SCATTERER}(b), the left and right surfaces
of the scatterer $A$ ($B$) are $a_{L}$ and $a_{R}$ ($b_{L}$ and $b_{R}$) and
the three leads sandwiching the scatterers are the semi-infinite left lead
$L,$ the middle lead $M,$ and the semi-infinite right lead $R$. For a local
excitation $|\Phi_{\mathrm{loc}}\rangle_{m_{0}}$ at $m_{0}\in C$, the total
scattering state inside $C$ is $|\Phi(m)\rangle|_{m\in C}%
=\mathbb{G}_{m,m_{0}}|\Phi_{\mathrm{loc}}\rangle_{m_{0}}$, which allows us to
identify $\mathbb{G}$ once the total scattering state has been obtained. For
$m_{0}\in A$, the local excitation first produces a zeroth-order partial wave
in $A$ and $M$. Next the right-going partial wave in $M$ undergoes multiple
reflections between $B$ and $A$ and finally evolves to a scattering state. For
$m_{0}\in B$, the local excitation first produces a zeroth-order partial wave
inside $B$ and $M$. Next the left-going partial wave in $M$ undergoes multiple
reflections and finally evolves to a scattering state. For $m_{0}\in M$, the
local excitation first produces a zeroth-order outgoing partial wave in $M$.
Next the left- and right-going partial waves each undergo multiple
reflections and evolve to a scattering state. For each case, the total
scattering state emanating from the local excitation is the sum of the
unscattered zeroth-order partial wave and the scattering state(s) emanating
from the scattered zeroth-order partial wave. Therefore, the key issue is to
calculate the scattering state $|\Psi(m)\rangle$ emanating from a right- or
left-going incident wave $|\Phi_{\mathrm{in}}(m)\rangle$ in the middle lead
through multiple reflections between $A$ and $B$.

In the middle lead, $|\Psi(m)\rangle$ is the sum of the right-going part
$|\Psi_{+}(m)\rangle=(\mathbf{P}_{+}^{(M)})^{m-b_{L}}|\Psi_{+}(b_{L})\rangle$
and the left-going part $|\Psi_{-}(m)\rangle=(\mathbf{P}_{-}^{(M)})^{m-a_{R}%
}|\Psi_{-}(a_{R})\rangle$. Inside the scatterer $A$, $|\Psi(m)\rangle$ is
equal to the scattering state emanating from the total incident wave
$|\Psi_{-}(a_{R})\rangle$ on $A$ [cf. Eq. (\ref{PHI_S2})]:%
\end{subequations}
\begin{equation}
|\Psi(m)\rangle|_{m\in A}=\mathbb{G}_{m,a_{R}}^{(A)}(\mathbf{g}^{(M)}%
)^{-1}|\Psi_{-}(a_{R})\rangle,
\end{equation}
Inside the scatterer $B$, $|\Psi(m)\rangle$ is equal to the scattering state
emanating from the total incident wave $|\Psi_{+}(b_{L})\rangle$ on $B$ [cf.
Eq. (\ref{PHI_S1})]:%
\begin{equation}
|\Psi(m)\rangle|_{m\in B}=\mathbb{G}_{m,b_{L}}^{(B)}(\mathbf{g}^{(M)}%
)^{-1}|\Psi_{+}(b_{L})\rangle.
\end{equation}
Therefore, the scattering state inside $C$ is completely determined by
$|\Psi_{+}(b_{L})\rangle$ and $|\Psi_{-}(a_{R})\rangle$. For brevity, we
introduce the reflection matrices
\begin{align}
\mathbb{R}_{B}  &  \equiv\mathbb{G}_{b_{L},b_{L}}^{(B)}(\mathbf{g}^{(M)}%
)^{-1}-\mathbf{I},\\
\mathbb{R}_{A}  &  \equiv\mathbb{G}_{a_{R},a_{R}}^{(A)}(\mathbf{g}^{(M)}%
)^{-1}-\mathbf{I}.
\end{align}
The former (latter) converts a right-going (left-going)\ incident wave on the
left (right) surface of $B$ ($A$) to a left-going (right-going)\ reflection
wave. To describe the multiple reflections between $A$ and $B$, we introduce
the following renormalized propagation matrices that incorporate multiple
reflections:%
\begin{align}
\mathbf{P}_{b_{L}\leftarrow a_{R}}  &  \equiv\lbrack1-(\mathbf{P}_{+}%
^{(M)})^{\Delta m}\mathbb{R}_{A}(\mathbf{P}_{-}^{(M)})^{-\Delta m}%
\mathbb{R}_{B}]^{-1}(\mathbf{P}_{+}^{(M)})^{\Delta m},\\
\mathbf{P}_{a_{R}\leftarrow b_{L}}  &  \equiv\lbrack1-(\mathbf{P}_{-}%
^{(M)})^{-\Delta m}\mathbb{R}_{B}(\mathbf{P}_{+}^{(M)})^{\Delta m}%
\mathbb{R}_{A}]^{-1}(\mathbf{P}_{-}^{(M)})^{-\Delta m},\\
\mathbf{P}_{a_{R}\leftarrow a_{R}}  &  \equiv(\mathbf{P}_{-}^{(M)})^{-\Delta
m}\mathbb{R}_{B}\mathbf{P}_{b_{L}\leftarrow a_{R}}=\mathbf{P}_{a_{R}\leftarrow
b_{L}}\mathbb{R}_{B}(\mathbf{P}_{+}^{(M)})^{\Delta m},\\
\mathbf{P}_{b_{L}\leftarrow b_{L}}  &  \equiv(\mathbf{P}_{+}^{(M)})^{\Delta
m}\mathbb{R}_{A}\mathbf{P}_{a_{R}\leftarrow b_{L}}=\mathbf{P}_{b_{L}\leftarrow
a_{R}}\mathbb{R}_{A}(\mathbf{P}_{-}^{(M)})^{-\Delta m},
\end{align}
where $\Delta m\equiv b_{L}-a_{R}$ is the distance between $A$ and $B$. For
example, the renormalized propagation matrix $\mathbf{P}_{b_{L}\leftarrow
a_{R}}$ from $a_{R}$ to $b_{L}$ is the sum of the free propagation term
$(\mathbf{P}_{+}^{(M)})^{\Delta m}$, the propagation term with two reflections
$(\mathbf{P}_{+}^{(M)})^{\Delta m}\mathbb{R}_{A}(\mathbf{P}_{-}^{(M)}%
)^{-\Delta m}\mathbb{R}_{B}(\mathbf{P}_{+}^{(M)})^{\Delta m}$, and so on.

Using the above notations, when $|\Phi_{\mathrm{in}}(m)\rangle$ is a
right-going incident wave on $B$, we have%
\begin{align}
|\Psi_{+}(b_{L})\rangle &  =(1+\mathbf{P}_{b_{L}\leftarrow b_{L}}%
\mathbb{R}_{B})|\Phi_{\mathrm{in}}(b_{L})\rangle,\\
|\Psi_{-}(a_{R})\rangle &  =\mathbf{P}_{a_{R}\leftarrow b_{L}}\mathbb{R}%
_{B}|\Phi_{\mathrm{in}}(b_{L})\rangle.
\end{align}
When $|\Phi_{\mathrm{in}}(m)\rangle$ is a left-going incident wave on $A$, we
have
\begin{align}
|\Psi_{+}(b_{L})\rangle &  =\mathbf{P}_{b_{L}\leftarrow a_{R}}\mathbb{R}%
_{A}|\Phi_{\mathrm{in}}(a_{R})\rangle,\\
|\Psi_{-}(a_{R})\rangle &  =(1+\mathbf{P}_{a_{R}\leftarrow a_{R}}%
\mathbb{R}_{A})|\Phi_{\mathrm{in}}(a_{R})\rangle.
\end{align}

Using the above results together with $|\Phi(m)\rangle|_{m\in C}%
=\mathbb{G}_{m,m_{0}}|\Phi_{\mathrm{loc}}\rangle_{m_{0}}$, we identify
\begin{widetext}
\begin{subequations}
\label{G_ALL}%
\begin{align}
\mathbb{G}_{m\in B,m_{0}\in A} &  =\mathbb{G}_{m,b_{L}}^{(B)}(\mathbf{g}%
^{(M)})^{-1}\mathbf{P}_{b_{L}\leftarrow a_{R}}\mathbb{G}_{a_{R},m_{0}}%
^{(A)},\label{GBA}\\
\mathbb{G}_{m\in A,m_{0}\in B} &  =\mathbb{G}_{m,a_{R}}^{(A)}(\mathbf{g}%
^{(M)})^{-1}\mathbf{P}_{a_{R}\leftarrow b_{L}}\mathbb{G}_{b_{L},m_{0}}%
^{(B)},\label{GAB}\\
\mathbb{G}_{m\in A,m_{0}\in A} &  =\mathbb{G}_{m,m_{0}}^{(A)}+\mathbb{G}%
_{m,a_{R}}^{(A)}(\mathbf{g}^{(M)})^{-1}\mathbf{P}_{a_{R}\leftarrow a_{R}%
}\mathbb{G}_{a_{R},m_{0}}^{(A)},\label{GAA}\\
\mathbb{G}_{m\in B,m_{0}\in B} &  =\mathbb{G}_{m,m_{0}}^{(B)}+\mathbb{G}%
_{m,b_{L}}^{(B)}(\mathbf{g}^{(M)})^{-1}\mathbf{P}_{b_{L}\leftarrow b_{L}%
}\mathbb{G}_{b_{L},m_{0}}^{(B)},\label{GBB}\\
\mathbb{G}_{m\in M,m_{0}\in A} &  =[(\mathbf{P}_{+}^{(M)})^{m-a_{R}%
}(1+\mathbb{R}_{A}\mathbf{P}_{a_{R}\leftarrow a_{R}})+(\mathbf{P}_{-}%
^{(M)})^{m-b_{L}}\mathbb{R}_{B}\mathbf{P}_{b_{L}\leftarrow a_{R}}%
]\mathbb{G}_{a_{R},m_{0}}^{(A)},\label{GMA}\\
\mathbb{G}_{m\in A,m_{0}\in M} &  =\mathbb{G}_{m,a_{R}}^{(A)}(\mathbf{g}%
^{(M)})^{-1}[(1+\mathbf{P}_{a_{R}\leftarrow a_{R}}\mathbb{R}_{A}%
)\mathbf{g}_{a_{R},m_{0}}^{(M)}+\mathbf{P}_{a_{R}\leftarrow b_{L}}%
\mathbb{R}_{B}\mathbf{g}_{b_{L},m_{0}}^{(M)}],\label{GAM}\\
\mathbb{G}_{m\in M,m_{0}\in B} &  =[(\mathbf{P}_{-}^{(M)})^{m-b_{L}%
}(1+\mathbb{R}_{B}\mathbf{P}_{b_{L}\leftarrow b_{L}})+(\mathbf{P}_{+}%
^{(M)})^{m-a_{R}}\mathbb{R}_{A}\mathbf{P}_{a_{R}\leftarrow b_{L}}%
]\mathbb{G}_{b_{L},m_{0}}^{(B)},\label{GMB}\\
\mathbb{G}_{m\in B,m_{0}\in M} &  =\mathbb{G}_{m,b_{L}}^{(B)}(\mathbf{g}%
^{(M)})^{-1}[(1+\mathbf{P}_{b_{L}\leftarrow b_{L}}\mathbb{R}_{B}%
)\mathbf{g}_{b_{L},m_{0}}^{(M)}+\mathbf{P}_{b_{L}\leftarrow a_{R}}%
\mathbb{R}_{A}\mathbf{g}_{a_{R},m_{0}}^{(M)}],\label{GBM}\\
\mathbb{G}_{m\in M,m_{0}\in M} &  =\mathbf{g}_{m,m_{0}}^{(M)}+(\mathbf{P}%
_{+}^{(M)})^{m-a_{R}}\mathbb{R}_{A}\mathbf{P}_{a_{R}\leftarrow b_{L}%
}\mathbb{R}_{B}\mathbf{g}_{b_{L},m_{0}}^{(M)}+(\mathbf{P}_{-}^{(M)})^{m-b_{L}%
}\mathbb{R}_{B}\mathbf{P}_{b_{L}\leftarrow a_{R}}\mathbb{R}_{A}\mathbf{g}%
_{a_{R},m_{0}}^{(M)}\label{GMM}\\
&  +(\mathbf{P}_{-}^{(M)})^{m-b_{L}}(\mathbb{R}_{B}+\mathbb{R}_{B}%
\mathbf{P}_{b_{L}\leftarrow b_{L}}\mathbb{R}_{B})\mathbf{g}_{b_{L},m_{0}%
}^{(M)}+(\mathbf{P}_{+}^{(M)})^{m-a_{R}}(\mathbb{R}_{A}+\mathbb{R}%
_{A}\mathbf{P}_{a_{R}\leftarrow a_{R}}\mathbb{R}_{A})\mathbf{g}_{a_{R},m_{0}%
}^{(M)}.\nonumber
\end{align}
\end{subequations}
\end{widetext}These equations can be interpreted in a physically transparent way. For
example, Eq. (\ref{GBA}) shows that the local excitation at $m_{0}\in A$
evolves to the scattering wave at $m\in B$ through the following steps: first
it is converted by $\mathbb{G}_{a_{R},m_{0}}^{(A)}$ to a zeroth-order partial
wave at $a_{R}$, next it undergoes renormalized propagation from $a_{R}$ to
$b_{L}$, and finally it is converted back to a local excitation and then to
the scattering wave at $m\in B$.

\subsubsection{Analytical construction rule for multiple scatterers}

\begin{figure}[t]
\includegraphics[width=\columnwidth,clip]{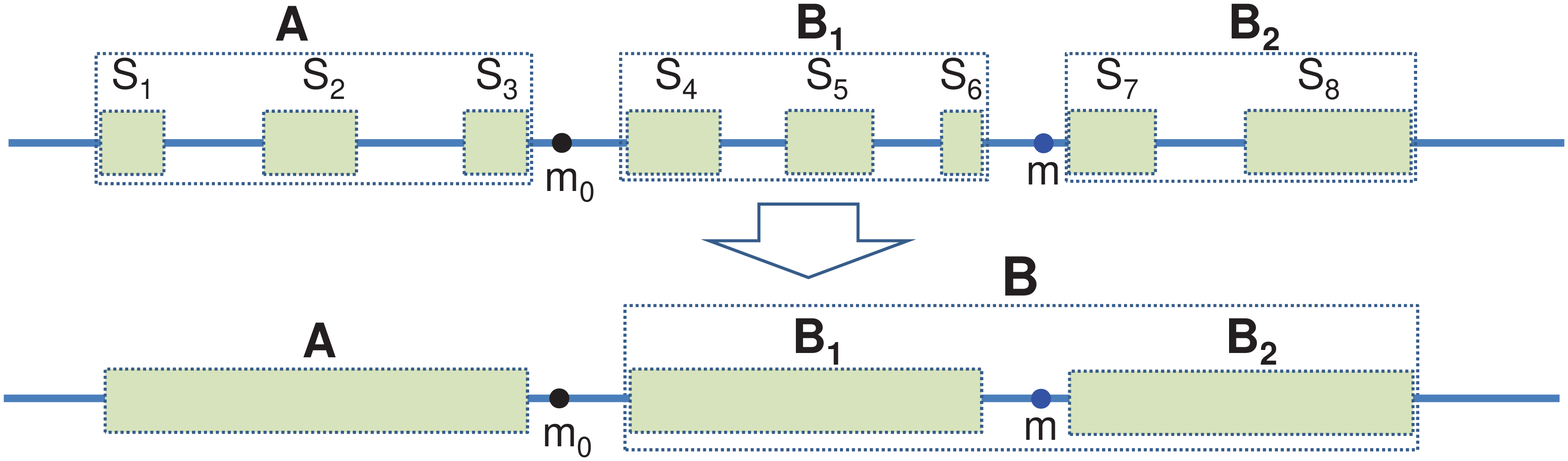}
 \caption{Green's
function $\mathbf{G}_{m,m_{0}}$ of an infinite system containing eight
scatterers, where\ $m_{0}$ and $m$ are both inside the leads or at the
surfaces of the scatterers.}%
\label{G_MULTI_EXAMPLE}%
\end{figure}

By repeatedly using Eq. (\ref{G_ALL}), the conversion matrix of a composite
scatterer can be obtained analytically as functions of the conversion matrices
of the constituent scatterers. In particular, the four surface elements of the
conversion matrix of the composite scatterer can be immediately obtained from
those of the constituent scatterers:
\begin{subequations}
\label{G_BOUNDARY}%
\begin{align}
\mathbb{G}_{b_{R},a_{L}}  &  =\mathbb{G}_{b_{R},b_{L}}^{(B)}(\mathbf{g}%
^{(M)})^{-1}\mathbf{P}_{b_{L}\leftarrow a_{R}}\mathbb{G}_{a_{R},a_{L}}%
^{(A)},\\
\mathbb{G}_{a_{L},b_{R}}  &  =\mathbb{G}_{a_{L},a_{R}}^{(A)}(\mathbf{g}%
^{(M)})^{-1}\mathbf{P}_{a_{R}\leftarrow b_{L}}\mathbb{G}_{b_{L},b_{R}}%
^{(B)},\\
\mathbb{G}_{a_{L},a_{L}}  &  =\mathbb{G}_{a_{L},a_{L}}^{(A)}+\mathbb{G}%
_{a_{L},a_{R}}^{(A)}(\mathbf{g}^{(M)})^{-1}\mathbf{P}_{a_{R}\leftarrow a_{R}%
}\mathbb{G}_{a_{R},a_{L}}^{(A)},\\
\mathbb{G}_{b_{R},b_{R}}  &  =\mathbb{G}_{b_{R},b_{R}}^{(B)}+\mathbb{G}%
_{b_{R},b_{L}}^{(B)}(\mathbf{g}^{(M)})^{-1}\mathbf{P}_{b_{L}\leftarrow b_{L}%
}\mathbb{G}_{b_{L},b_{R}}^{(B)},
\end{align}
while the latter can be calculated through recursive techniques [Eqs.
(\ref{RECURSIVE1}) and (\ref{RECURSIVE2})].

For example, let us consider an infinite layered system with eight scatterers
$S_{1},S_{2},\cdots,S_{8}$ and calculate its GF $\mathbf{G}_{m,m_{0}}$.
For simplicity we assume that $m_{0}$ and $m$ are both inside the leads or at the surfaces of the scatterers (see Fig. 6), so that $\mathbf{G}_{m,m_{0}}$ is just the $(m,m_{0})$ element of the conversion matrix $\mathbb{G}$ of the composite scatterer $(S_{1}+S_{2}+\cdots+S_{8})$, i.e., $\mathbf{G}_{m,m_{0}}=\mathbb{G}_{m,m_{0}}$, and $\mathbb{G}_{m,m_{0}}$ is completely
determined by the surface elements of $\mathbb{G}^{(S_{1})},\cdots
,\mathbb{G}^{(S_{8})}$, which are readily obtained through recursive
techniques. First, we use Eq. (\ref{G_BOUNDARY}) to calculate the surface
elements of $\mathbb{G}^{(A)}$, $\mathbb{G}^{(B_{1})}$, and $\mathbb{G}%
^{(B_{2})}$ for the three composite scatterers $A\equiv(S_{1}+S_{2}+S_{3})$,
$B_{1}\equiv(S_{4}+S_{5}+S_{6})$, and $B_{2}\equiv(S_{7}+S_{8})$. Next we
regard $(B_{1}+B_{2})$ as a composite scatterer $B$ and calculate
$\mathbb{G}_{m,b_{L}}^{(B)}$ ($b_{L}$ is the left surface of $B$) from Eq.
(\ref{GMA}). Now the entire system contains two scatterers\ $A$ and $B$; thus
$\mathbb{G}_{m,m_{0}}$ can be obtained from Eq. (\ref{GBM}).

\section{Inverse of Fisher-Lee relation}

In the previous section, we have developed a physically transparent and
numerically efficient way to calculate the GF of a general layered system.
There the GF is expressed as a matrix, i.e., in terms of the propagation
matrices $\mathbf{P}_{\pm}$ and conversion matrix $\mathbf{g}$ of the leads
and the conversion matrices $\mathbb{G}$ of the scatterers. In this section,
we give further physical insight into our GF approach by expressing the GF analytically in
terms of a generalized scattering matrix, which describes the scattering of
both traveling states and evanescent states. This could be regarded as the
inverse of the well-known Fisher-Lee relations
\cite{FisherPRB1981,StoneIJRD1988,BarangerPRB1989,SolsAP1992,SanvitoPRB1999,KhomyakovPRB2005,WimmerThesis2009} (see Fig. 1). The key is to express the conversion matrix $\mathbb{G}$ in terms of a
generalized scattering matrix.

\subsection{Generalized scattering matrix}

Let us consider an infinite system consisting of a single scatterer (with the
left surface at $m_{L}$ and the right surface at $m_{R}$)\ connected to two
semi-infinite leads $L$ and $R$ (see Fig. \ref{G_SINGLE}). For a right-going
eigenmode $|\Phi_{\mathrm{in}}(m)\rangle=e^{ik_{+,\alpha}^{(L)}(m-m_{L})a_{L}%
}|\Phi_{+,\alpha}^{(L)}\rangle$ incident on the scatterer from the left lead,
the resulting scattering state $|\Psi(m)\rangle$ follows from Eqs.
(\ref{PHI_S1})-(\ref{PHI_MR}). Using Eq. (\ref{P_EXPAND}) for the propagation
matrix, we obtain
\end{subequations}
\begin{align}
|\Psi(m)\rangle|_{m\geq m_{R}}  &  =\sum_{\beta}e^{ik_{+,\beta}^{(R)}%
(m-m_{R})a_{R}}|\Phi_{+,\beta}^{(R)}\rangle\mathcal{S}_{\beta,\alpha}^{(RL)}\\
|\Psi(m)\rangle|_{m\leq m_{L}}  &  =|\Phi_{\mathrm{in}}(m)\rangle+\sum_{\beta
}e^{ik_{-,\beta}^{(L)}(m-m_{L})a_{L}}|\Phi_{-,\beta}^{(L)}\rangle
\mathcal{S}_{\beta,\alpha}^{(LL)},
\end{align}
where
\begin{equation}
\mathcal{S}_{\beta,\alpha}^{(RL)}\equiv\langle\phi_{+,\beta}^{(R)}%
|\mathbb{G}_{m_{R},m_{L}}(\mathbf{g}^{(L)})^{-1}|\Phi_{+,\alpha}^{(L)}%
\rangle\label{SRL}%
\end{equation}
is a generalized transmission amplitude from $|\Phi_{+,\alpha}^{(L)}\rangle$
at the left surface of the scatterer into $|\Phi_{+,\beta}^{(R)}\rangle$ at
the right surface of the scatterer, and
\begin{equation}
\mathcal{S}_{\beta,\alpha}^{(LL)}\equiv\langle\phi_{-,\beta}^{(L)}%
|[\mathbb{G}_{m_{L},m_{L}}(\mathbf{g}^{(L)})^{-1}-\mathbf{I}]|\Phi_{+,\alpha
}^{(L)}\rangle\label{SLL}%
\end{equation}
is a generalized reflection amplitude from $|\Phi_{+,\alpha}^{(L)}\rangle$
into $|\Phi_{-,\beta}^{(L)}\rangle$ at the left surface of the scatterer.
Similarly, for a left-going incident wave $|\Phi_{\mathrm{in}}(m)\rangle
=e^{ik_{-,\alpha}^{(R)}(m-m_{R})a_{R}}|\Phi_{-,\alpha}^{(R)}\rangle$ in the
right lead, the resulting scattering state $|\Psi(m)\rangle$ follows from Eqs.
(\ref{PHI_S2})-(\ref{PHI_ML}) as
\begin{align}
|\Psi(m)\rangle|_{m\leq m_{L}}  &  =\sum_{\beta}e^{ik_{-,\beta}^{(L)}%
(m-m_{L})a_{L}}|\Phi_{-,\beta}^{(L)}\rangle\mathcal{S}_{\beta,\alpha}%
^{(LR)},\\
|\Psi(m)\rangle|_{m\geq m_{R}}  &  =|\Phi_{\mathrm{in}}(m)\rangle+\sum_{\beta
}e^{ik_{+,\beta}^{(R)}(m-m_{R})a_{R}}\mathcal{S}_{\beta,\alpha}^{(RR)}%
|\Phi_{+,\beta}^{(R)}\rangle,
\end{align}
where
\begin{equation}
\mathcal{S}_{\beta,\alpha}^{(LR)}\equiv\langle\phi_{-,\beta}^{(L)}%
|\mathbb{G}_{m_{L},m_{R}}(\mathbf{g}^{(R)})^{-1}|\Phi_{-,\alpha}^{(R)}%
\rangle\label{SLR}%
\end{equation}
is a generalized transmission amplitude from $|\Phi_{-,\alpha}^{(R)}\rangle$
at the right surface of the scatterer into $|\Phi_{-,\beta}^{(L)}\rangle$ at
the left surface of the scatterer, and
\begin{equation}
\mathcal{S}_{\beta,\alpha}^{(RR)}\equiv\langle\phi_{+,\beta}^{(R)}%
|[\mathbb{G}_{m_{R},m_{R}}(\mathbf{g}^{(R)})^{-1}-\mathbf{I}]|\Phi_{-,\alpha
}^{(R)}\rangle\label{SRR}%
\end{equation}
is a generalized reflection amplitude from $|\Phi_{-,\alpha}^{(R)}\rangle$
into $|\Phi_{+,\beta}^{(R)}\rangle$ at the right surface of the scatterer.

Equations (\ref{SRL})-(\ref{SRR}) define a generalized scattering matrix
$\mathcal{S}(E)$ and express it in terms of the surface elements of the
conversion matrix. They were first derived in the wave function mode matching
approach \cite{AndoPRB1991,NikolicPRB1994,XiaPRB2006} and its connection to
the GF approach was established later \cite{KhomyakovPRB2005}. They are valid
for both traveling modes and evanescent modes. In our GF approach, these expressions have clear physical
meanings. Taking $\mathcal{S}_{\beta,\alpha}^{(RL)}$ as an example,
$\mathbb{G}_{m_{R},m_{L}}(\mathbf{g}^{(L)})^{-1}$ converts the incident
eigenmode $|\Phi_{+,\alpha}^{(L)}\rangle$ at the left surface of the scatterer
back to a local excitation and then to the scattering wave at the right
surface of the scatterer. Then the dual vector $\langle\phi_{+,\beta}^{(R)}|$
projects the scattering wave onto the eigenmode $|\Phi_{+,\beta}^{(R)}\rangle
$ [see Eqs. (24) and (25)]. The transmission and reflection amplitudes of the unitary scattering matrix
connecting two traveling eigenmodes are obtained by normalizing with respect
to the current \cite{KhomyakovPRB2005}:
\begin{equation}
S_{\beta,\alpha}^{(q,p)}|_{\alpha,\beta\in\mathrm{traveling}}=\sqrt
{\frac{|v_{s_{\mathrm{out}},\beta}^{(q)}| /a_{q}}{|v_{s_{\mathrm{in}},\alpha}^{(p)}|/q_{p}}}\mathcal{S}%
_{\beta,\alpha}^{(q,p)},
\end{equation}
where $s_{\mathrm{in}}=s_{\mathrm{out}}=+$ for $(q,p)=(R,L)$; $s_{\mathrm{in}}=s_{\mathrm{out}}=-$ for $(q,p)=(L,R)$;
$s_{\mathrm{in}}=+$, $s_{\mathrm{out}}=-$ for $(q,p)=(L,L)$; and $s_{\mathrm{in}}=-$, $s_{\mathrm{out}}=+$ for $(q,p)=(R,R)$,
thus Eqs. (\ref{SRL}) and (\ref{SLR}) lead to Eqs.
(\ref{SUNITARY_RL}) and (\ref{SUNITARY_LR}), respectively.

\subsection{Inverse of Fisher-Lee relations}

The inverse of Eqs. (\ref{SRL})-(\ref{SRR}) gives the surface elements of
$\mathbb{G}(E)$ in terms of the generalized scattering matrix $\mathcal{S}%
(E)$:
\begin{subequations}
\label{G_BOUNDARY_EXP}%
\begin{align}
\mathbb{G}_{m_{R},m_{L}}  &  =\sum_{\alpha}\left(  \sum_{\beta}\mathcal{S}%
_{\beta,\alpha}^{(RL)}|\Phi_{+,\beta}^{(R)}\rangle\right)  \langle
\phi_{+,\alpha}^{(L)}|\mathbf{g}^{(L)},\label{GB_RL_EXP}\\
\mathbb{G}_{m_{L},m_{R}}  &  =\sum_{\alpha}\left(  \sum_{\beta}\mathcal{S}%
_{\beta,\alpha}^{(LR)}|\Phi_{-,\beta}^{(L)}\rangle\right)  \langle
\phi_{-,\alpha}^{(R)}|\mathbf{g}^{(R)},\label{GB_LR_EXP}\\
\mathbb{G}_{m_{L},m_{L}}  &  =\sum_{\alpha}\left(  |\Phi_{+,\alpha}%
^{(L)}\rangle+\sum_{\beta}\mathcal{S}_{\beta,\alpha}^{(LL)}|\Phi_{-,\beta
}^{(L)}\rangle\right)  \langle\phi_{+,\alpha}^{(L)}|\mathbf{g}^{(L)}%
,\label{GB_LL_EXP}\\
\mathbb{G}_{m_{R},m_{R}}  &  =\sum_{\alpha}\left(  |\Phi_{-,\alpha}%
^{(R)}\rangle+\sum_{\beta}\mathcal{S}_{\beta,\alpha}^{(RR)}|\Phi_{+,\beta
}^{(R)}\rangle\right)  \langle\phi_{-,\alpha}^{(R)}|\mathbf{g}^{(R)}.
\label{GB_RR_EXP}%
\end{align}
These expressions have very clear physical interpretations. For example, Eq.
(\ref{GB_RL_EXP}) shows that a local excitation $|\Phi_{\mathrm{loc}}%
\rangle_{m_{L}}$ at the left surface of the scatterer evolves to the
scattering wave at the right surface of the scatterer through two steps.
First, it evolves to a partial wave $\mathbf{g}^{(L)}|\Phi_{\mathrm{loc}%
}\rangle_{m_{L}}$, then it is expanded into linear combinations of right-going
eigenmodes $\sum_{\alpha}|\Phi_{+,\alpha}^{(L)}\rangle\langle\phi_{+,\alpha
}^{(L)}|\mathbf{g}^{(L)}|\Phi_{\mathrm{loc}}\rangle_{m_{L}}$ and each
eigenmode transmits across the scatterer to its right surface as
\end{subequations}
\begin{equation}
|\Phi_{+,\alpha}^{(L)}\rangle\rightarrow\sum_{\beta}\mathcal{S}_{\beta,\alpha
}^{(RL)}|\Phi_{+,\beta}^{(R)}\rangle.
\end{equation}
Similarly, Eq. (\ref{GB_LL_EXP}) shows that a local excitation at the left
surface of the scatterer first evolves to a partial wave $\mathbf{g}%
^{(L)}|\Phi_{\mathrm{loc}}\rangle_{m_{L}}$, then it is expanded as
$\sum_{\alpha}|\Phi_{+,\alpha}^{(L)}\rangle\langle\phi_{+,\alpha}%
^{(L)}|\mathbf{g}^{(L)}|\Phi_{\mathrm{loc}}\rangle_{m_{L}}$ and each eigenmode
evolves to a scattering wave:\
\begin{equation}
|\Phi_{+,\alpha}^{(L)}\rangle\rightarrow|\Phi_{+,\alpha}^{(L)}\rangle
+\sum_{\beta}|\Phi_{-,\beta}^{(L)}\rangle\mathcal{S}_{\beta,\alpha}^{(LL)},
\end{equation}
which consists of the incident wave and the reflection wave.

Next, we can express other blocks of the GF, i.e., Eqs. (\ref{G_ALL_S}%
)-(\ref{G_ALL_R}) with $m,m_{0}$ inside the leads, in terms of the generalized
scattering matrix:
\begin{subequations}
\label{G_ALL_EXP}%
\begin{eqnarray}
\mathbf{G}_{m\in R,m_{0}\in L}&=&\sum_{\alpha}\left(  \sum_{\beta
}e^{ik_{+,\beta}^{(R)}(m-m_{R})a}|\Phi_{+,\beta}^{(R)}\rangle\mathcal{S}%
_{\beta,\alpha}^{(RL)}\right) \notag
\\ &&\times ~ e^{ik_{+,\alpha}^{(L)}(m_{L}-m_{0})a}%
\langle\phi_{+,\alpha}^{(L)}|\mathbf{g}^{(L)},\label{GRL_EXP}\\
\mathbf{G}_{m\in L,m_{0}\in R} &=&\sum_{\alpha}\left(  \sum_{\beta
}e^{ik_{-,\beta}^{(L)}(m-m_{L})a}|\Phi_{-,\beta}^{(L)}\rangle\mathcal{S}%
_{\beta,\alpha}^{(LR)}\right) \notag
\\&& \times ~ e^{ik_{-,\alpha}^{(R)}(m_{R}-m_{0})a}%
\langle\phi_{-,\alpha}^{(R)}|\mathbf{g}^{(R)},\label{GLR_EXP}\\
\mathbf{G}_{m\in L,m_{0}\in L} &=&\mathbf{g}_{m,m_{0}}^{(L)}+\sum_{\alpha
}\left(  \sum_{\beta}e^{ik_{-,\beta}^{(L)}(m-m_{L})a}|\Phi_{-,\beta}%
^{(L)}\rangle\mathcal{S}_{\beta,\alpha}^{(LL)}\right)  \notag
\\&& \times ~ e^{ik_{+,\alpha}%
^{(L)}(m_{L}-m_{0})a}\langle\phi_{+,\alpha}^{(L)}|\mathbf{g}^{(L)}%
,\label{GLL_EXP}\\
\mathbf{G}_{m\in R,m_{0}\in R} &=&\mathbf{g}_{m,m_{0}}^{(R)}+\sum_{\alpha
}\left(  \sum_{\beta}e^{ik_{+,\beta}^{(R)}(m-m_{R})a}|\Phi_{+,\beta}%
^{(R)}\rangle\mathcal{S}_{\beta,\alpha}^{(RR)}\right) \notag
\\&& \times ~ e^{ik_{-,\alpha}%
^{(R)}(m_{R}-m_{0})a}\langle\phi_{-,\alpha}^{(R)}|\mathbf{g}^{(R)}%
.\label{GRR_EXP}%
\end{eqnarray}
\end{subequations}
The above expressions have clear physical meanings. For example,
Eq. (\ref{GRL_EXP}) shows that a local excitation $|\Phi_{\mathrm{loc}}%
\rangle_{m_{0}}$ in the left lead first evolves to a partial wave
$\mathbf{g}^{(L)}|\Phi_{\mathrm{loc}}\rangle_{m_{0}}$ and then propagates
freely to the left surface of the scatterer as $\sum_{\alpha}e^{ik_{+,\alpha
}^{(L)}(m_{L}-m_{0})a}|\Phi_{+,\alpha}^{(L)}\rangle\langle\phi_{+,\alpha
}^{(L)}|\mathbf{g}^{(L)}|\Phi_{\mathrm{loc}}\rangle_{m_{0}}$. Finally each
eigenmode $|\Phi_{+,\alpha}^{(L)}\rangle$ evolves to a transmission wave:
\begin{equation}
|\Phi_{+,\alpha}^{(L)}\rangle\rightarrow\sum_{\beta}e^{ik_{+,\beta}%
^{(R)}(m-m_{R})a_{R}}|\Phi_{+,\beta}^{(R)}\rangle\mathcal{S}_{\beta,\alpha
}^{(RL)}.
\end{equation}
As another example, Eq. (\ref{GLL_EXP}) shows that $\mathbf{G}_{m\in
L,m_{0}\in L}$ is the sum of the free GF $\mathbf{g}_{m,m_{0}}^{(L)}$ and the
reflection wave contribution, which emerges as follows: the local excitation
$|\Phi_{\mathrm{loc}}\rangle_{m_{0}}$ in the left lead first evolves to a
partial wave $\mathbf{g}^{(L)}|\Phi_{\mathrm{loc}}\rangle_{m_{0}}$ and then
propagates freely to the left surface of the scatterer as $\sum_{\alpha
}e^{ik_{+,\alpha}^{(L)}(m_{L}-m_{0})a}|\Phi_{+,\alpha}^{(L)}\rangle\langle
\phi_{+,\alpha}^{(L)}|\mathbf{g}^{(L)}|\Phi_{\mathrm{loc}}\rangle_{m_{0}}$.
Finally, each mode evolves to a reflection wave:
\begin{equation}
|\Phi_{+,\alpha}^{(L)}\rangle\rightarrow\sum_{\beta}e^{ik_{-,\beta}%
^{(L)}(m-m_{L})a_{L}}|\Phi_{-,\beta}^{(L)}\rangle\mathcal{S}_{\beta,\alpha
}^{(LL)}.
\end{equation}
Equation (\ref{G_ALL_EXP}) not only allows us to construct the GF from the
generalized scattering matrix, but also reveals the contribution of each
individual scattering channels to the GF.

\subsection{On-shell spectral expansion}

The Fisher-Lee relations
\cite{FisherPRB1981,StoneIJRD1988,BarangerPRB1989,SolsAP1992,SanvitoPRB1999,KhomyakovPRB2005,WimmerThesis2009}
and its inverse [Eq. (\ref{G_ALL_EXP})] provide a complete one-to-one
correspondence between the lattice GF approach and the wave function
mode matching approach
\cite{AndoPRB1991,NikolicPRB1994,KhomyakovPRB2005,XiaPRB2006} to mesoscopic
quantum transport (see Fig. 1). Next we show that we can construct the GF $\mathbf{G}(E)$
analytically in terms of a few scattering states on the energy shell $E$. Since
the latter can be readily obtained from standard textbook technique and
approximation methods (such as the WKB approximation), this provides a
convenient way to obtain the GF.

Let us introduce $2M$ \textit{advanced} eigenmodes $\{\tilde{k}_{s,\alpha
},|\tilde{\Phi}_{s,\alpha}\rangle\}$ of each lead \cite{KhomyakovPRB2005},%
\begin{equation}%
\begin{array}
[c]{ccc}%
\tilde{k}_{s,\alpha}\equiv k_{s,\alpha}, & |\tilde{\Phi}_{s,\alpha}%
\rangle\equiv|\Phi_{s,\alpha}\rangle\  & (\alpha\in\text{\textrm{evanescent}%
}),\\
\tilde{k}_{s,\alpha}\equiv k_{-s,\alpha}, & |\tilde{\Phi}_{s,\alpha}%
\rangle\equiv|\Phi_{-s,\alpha}\rangle & (\alpha\in\text{\textrm{traveling}}),
\end{array}
\end{equation}
and their dual vectors:%
\begin{equation}%
\begin{bmatrix}
\langle\tilde{\phi}_{s,1}|\\
\vdots\\
\langle\tilde{\phi}_{s,M}|
\end{bmatrix}
\equiv\mathbf{\tilde{U}}_{s}^{-1},
\end{equation}
where $\mathbf{\tilde{U}}_{s}\equiv\lbrack|\tilde{\Phi}_{s,1}\rangle
,\cdots,|\tilde{\Phi}_{s,M}\rangle]$. The advanced eigenvectors $\{|\tilde
{\Phi}_{s,\alpha}\rangle\}$ and dual vectors $\{|\tilde{\phi}_{s,\alpha
}\rangle\}$ obey exactly the same orthonormal and completeness relations as
their retarded counterpart [Eqs. (\ref{ORTHO}) and (\ref{COMPLETE})]. Using
\cite{KhomyakovPRB2005}%
\begin{equation}
\mathbf{g}^{-1}=\sum_{\alpha}|\tilde{\phi}_{-,\alpha}\rangle\frac
{iv_{+,\alpha}}{a}\langle\phi_{+,\alpha}|=\sum_{\alpha}|\tilde{\phi}%
_{+,\alpha}\rangle\frac{-iv_{-,\alpha}}{a}\langle\phi_{-,\alpha}|,
\label{GINV_EXPANSION}%
\end{equation}
and the completeness relations Eq. (\ref{COMPLETE}), we obtain%
\begin{align}
\mathbf{g}  &  =\sum_{\alpha}|\Phi_{+,\alpha}\rangle\frac{a}{iv_{+,\alpha}%
}\langle\tilde{\Phi}_{-,\alpha}|=\sum_{\alpha}|\Phi_{-,\alpha}\rangle\frac
{a}{-iv_{-,\alpha}}\langle\tilde{\Phi}_{+,\alpha}|\label{G_EXPANSION}\\
&  \longrightarrow\sum_{\alpha}|\Phi_{+,\alpha}\rangle\frac{a}{iv_{+,\alpha}%
}\langle\Phi_{+,\alpha}|=\sum_{\alpha}|\Phi_{-,\alpha}\rangle\frac
{a}{-iv_{-,\alpha}}\langle\Phi_{-,\alpha}|,
\end{align}
where the second line holds when all the eigenmodes are traveling modes. Here
$a$ is the unit cell spacing of the lead and $v_{s,\alpha}$ is the generalized
group velocity of the eigenmode $(s,\alpha)$:\ for a traveling mode, it equals
the group velocity [Eq. (\ref{VELOCITY})]; for an evanescent mode, it is
defined as
\begin{equation}
v_{+,\alpha}=v_{-,\alpha}^{\ast}=-ia\langle\Phi_{-,\alpha}|(e^{ik_{-,\alpha}%
a})^{\ast}\mathbf{t}^{\dagger}-e^{ik_{+,\alpha}a}\mathbf{t}|\Phi_{+,\alpha
}\rangle. \label{VSA}%
\end{equation}
Note that for an evanescent (traveling) mode, $v_{s,\alpha}$ depends (does not
depend) on the choice of the phases of the eigenvectors $\{|\Phi_{s,\alpha
}\rangle\}$.

To express the GF in terms of on-shell scattering states, we introduce the
eigenmode wave functions
\begin{subequations}
\label{BULK_MODE_DEF}%
\begin{align}
|\Phi_{+,\alpha}(m)\rangle &  \equiv\left\{
\begin{array}
[c]{ll}%
e^{ik_{+,\alpha}(m-m_{0})a}|\Phi_{+,\alpha}\rangle & (m\geq m_{0}),\\
0 & (m<m_{0}),
\end{array}
\right. \\
|\Phi_{-,\alpha}(m)\rangle &  \equiv\left\{
\begin{array}
[c]{ll}%
0 & (m\geq m_{0}),\\
e^{ik_{-,\alpha}(m-m_{0})a}|\Phi_{-,\alpha}\rangle & (m<m_{0}),
\end{array}
\right.
\end{align}
for the lead in which $m_{0}$ locates. Note that so-defined $|\Phi_{s,\alpha
}(m)\rangle$ depends on $m_{0}$, which is regarded as fixed and hence omitted
for brevity. If there were no scatterers, then $\mathbf{G}_{m,m_{0}}$ would
coincide with the free GF of this lead [Eq. (\ref{G_INFINITE})], which can be
written as%
\end{subequations}
\begin{equation}
\mathbf{g}_{m,m_{0}}=a%
{\displaystyle\sum\limits_{\alpha}}
\left(  \frac{|\Phi_{+,\alpha}(m)\rangle\langle\tilde{\Phi}_{-,\alpha}%
|}{iv_{+,\alpha}}+\frac{|\Phi_{-,\alpha}(m)\rangle\langle\tilde{\Phi
}_{+,\alpha}|}{-iv_{-,\alpha}}\right)  . \label{GFREE_SPECTRAL_EXP}%
\end{equation}
Due to the presence of scatterers, each eigenmode $|\Phi_{s,\alpha}(m)\rangle$
evolves to a corresponding scattering state $|\Psi_{s,\alpha}(m)\rangle$, so
replacing $|\Phi_{s,\alpha}(m)\rangle$ in Eq. (\ref{GFREE_SPECTRAL_EXP}) with
$|\Psi_{s,\alpha}(m)\rangle$ gives the GF:%
\begin{equation}
\mathbf{G}_{m,m_{0}}=a%
{\displaystyle\sum\limits_{\alpha}}
\left(  \frac{|\Psi_{+,\alpha}(m)\rangle\langle\tilde{\Phi}_{-,\alpha}%
|}{iv_{+,\alpha}}+\frac{|\Psi_{-,\alpha}(m)\rangle\langle\tilde{\Phi
}_{+,\alpha}|}{-iv_{-,\alpha}}\right)  . \label{G_SPECTRAL_EXP2}%
\end{equation}
Since the total scattering state $|\Psi_{s,\alpha}(m)\rangle$ can be
decomposed into the sum of the incident wave $|\Phi_{s,\alpha}(m)\rangle$
(which vanishes outside the lead in which $m_{0}$ locates) and the outgoing
scattering wave $|\Psi_{s,\alpha}^{(\mathrm{out})}(m)\rangle\equiv
|\Psi_{s,\alpha}(m)\rangle-|\Phi_{s,\alpha}(m)\rangle$, Eq.
(\ref{G_SPECTRAL_EXP2}) can also be written as
\begin{equation}
\mathbf{G}_{m,m_{0}}=\mathbf{g}_{m,m_{0}}+a%
{\displaystyle\sum\limits_{\alpha}}
\left(  \frac{|\Psi_{+,\alpha}^{(\mathrm{out})}(m)\rangle\langle\tilde{\Phi
}_{-,\alpha}|}{iv_{+,\alpha}}+\frac{|\Psi_{-,\alpha}^{(\mathrm{out}%
)}(m)\rangle\langle\tilde{\Phi}_{+,\alpha}|}{-iv_{-,\alpha}}\right)  ,
\label{G_SPECTRAL_EXP3}%
\end{equation}
where $\mathbf{g}_{m,m_{0}}$ is nonzero only for $m$ in the same lead as
$m_{0}$.

Equation (\ref{G_SPECTRAL_EXP2}) or (\ref{G_SPECTRAL_EXP3}) shows that the GF
$\mathbf{G}_{m,m_{0}}(E)$ is essentially certain scattering states $\{|\Psi_{s,\alpha}(m)\rangle\}$ (on the energy shell $E$) that obey
outgoing boundary conditions; i.e., they emanate from the on-shell eigenmodes
$|\Phi_{s,\alpha}(m)\rangle$ of the lead in which the local excitation occurs.
Compared with the standard spectral expansion in classic textbook on quantum
mechanics \cite{SakuraiBook1994,GriffithsBook1995,CohenBook2005} that
expresses the GF in terms of all the eigenstates (both on-shell eigenstates
and off-shell ones) of the system, Eq. (\ref{G_SPECTRAL_EXP2}) or
(\ref{G_SPECTRAL_EXP3}) deepens our physical understanding about the GF and
allows analytical reconstruction of the GF from a few scattering states of the system.

As an example, let us consider an infinite system containing a single scatterer
(Fig. \ref{G_SINGLE}). For $m_{0}$ in the left lead, the left-going eigenmode
$|\Phi_{-,\alpha}^{(L)}(m)\rangle$ is not scattered, so $|\Psi_{-,\alpha
}^{(L,\mathrm{out})}(m)\rangle=0$, while the right-going eigenmode
$|\Phi_{+,\alpha}^{(L)}(m)\rangle$ produces an outgoing scattering wave%
\begin{align}
|\Psi_{+,\alpha}^{(L,\mathrm{out})}(m)\rangle|_{m\in L}  &  =\left(
\sum_{\beta}e^{ik_{-,\beta}^{(L)}(m-m_{L})a}|\Phi_{-,\beta}^{(L)}%
\rangle\mathcal{S}_{\beta,\alpha}^{(LL)}\right)  e^{ik_{+,\alpha}^{(L)}%
(m_{L}-m_{0})a},\\
|\Psi_{+,\alpha}^{(L,\mathrm{out})}(m)\rangle|_{m\in R}  &  =\left(
\sum_{\beta}e^{ik_{+,\beta}^{(R)}(m-m_{R})a}|\Phi_{+,\beta}^{(R)}%
\rangle\mathcal{S}_{\beta,\alpha}^{(RL)}\right)  e^{ik_{+,\alpha}^{(L)}%
(m_{L}-m_{0})a}.
\end{align}
Substituting into Eq. (\ref{G_SPECTRAL_EXP3}) gives Eqs. (\ref{GRL_EXP}) and
(\ref{GLL_EXP}). Similarly, for $m_{0}$ in the right lead, the right-going
eigenmode $|\Phi_{+,\alpha}^{(R)}(m)\rangle$ is not scattered, so
$|\Psi_{+,\alpha}^{(R,\mathrm{out})}(m)\rangle=0$, while the left-going
eigenmode $|\Phi_{-,\alpha}^{(R)}(m)\rangle$ produces the outgoing wave
\begin{align}
|\Psi_{-,\alpha}^{(R,\mathrm{out})}(m)\rangle|_{m\in L}  &  =\left(
\sum_{\beta}e^{ik_{-,\beta}^{(L)}(m-m_{L})a}|\Phi_{-,\beta}^{(L)}%
\rangle\mathcal{S}_{\beta,\alpha}^{(LR)}\right)  e^{ik_{-,\alpha}^{(R)}%
(m_{R}-m_{0})\ a},\\
|\Psi_{-,\alpha}^{(R,\mathrm{out})}(m)\rangle|_{m\in R}  &  =\left(
\sum_{\beta}e^{ik_{+,\beta}^{(R)}(m-m_{R})a}|\Phi_{+,\beta}^{(R)}%
\rangle\mathcal{S}_{\beta,\alpha}^{(RR)}\right)  e^{ik_{-,\alpha}^{(R)}%
(m_{R}-m_{0})a}.
\end{align}
Substituting them into Eq. (\ref{G_SPECTRAL_EXP3}) gives Eqs. (\ref{GLR_EXP})
and (\ref{GRR_EXP}).

\section{Example and applications}

Here we first exemplify our general results in a 1D chain and then apply the
formalism to graphene \textit{p-n} junctions described by the tight-binding model.

\subsection{Simple example:\ 1D chain}

We consider a 1D chain characterized by one basis state in each unit cell, the
unit cell Hamiltonian $\mathbf{h}=\varepsilon_{0}$, and the nearest-neighbor
hopping $\mathbf{t}=-t<0$. For a given wave vector $k$, Eq. (\ref{MEQ1}) can
be solved to yield the energy $E(k)=\varepsilon_{0}-2t\cos(ka)$, which is real
in two cases: (1) $k\in\mathbb{R}$; (2) $k=i\kappa$ or $k=\pi/a+i\kappa$ with
$\kappa\in\mathbb{R}$. The former gives the real energy band$,$ while the
latter gives the complex energy bands. For specificity we consider the energy
$E\in\lbrack\varepsilon_{0}-2t,\varepsilon_{0}+2t]$, so there is one
right-going traveling eigenmode $k_{+}=k$ with group velocity $v=2at\sin
(ka)>0$ and one left-going traveling eigenmode $k_{-}=-k$ with group velocity
$-v$, where $k$ is the positive solution to $E=E(k)$. The eigenvectors of the
eigenmodes and dual vectors are $\Phi_{\pm}=\phi_{\pm}=1$. The propagation
matrices are $\mathbf{P}_{\pm}=e^{\pm ika}$, the conversion matrix of the lead
is%
\begin{equation}
\mathbf{g}=\frac{1}{2it\sin(ka)}=\frac{1}{iv/a},
\end{equation}
and the free GF of the lead is
\begin{equation}
\mathbf{g}_{m,m_{0}}=\frac{e^{ik|m-m_{0}|a}}{iv/a}.
\end{equation}

When the on-site energy of the unit cell $m_{1}$ of the infinite 1D chain is
replaced by $\varepsilon_{0}+\delta$, the unit cell $m_{1}$ becomes a
scatterer characterized the conversion matrix%
\begin{equation}
\mathbb{G}=\frac{1}{iv/a-\delta}, \label{CM_EXAMPLE}%
\end{equation}
or equivalently the transmission amplitude $\mathcal{T}=\mathbb{G}%
\mathbf{g}^{-1}=(iv/a)/(iv/a-\delta)$ and reflection amplitude $\mathcal{R}%
=\mathbb{G}\mathbf{g}^{-1}-1=\delta/(iv/a-\delta)$. The GFs of the entire
system are given by Eqs. (\ref{G_ALL_S}), (\ref{G_ALL_L}), and (\ref{G_ALL_R})
as%
\begin{align}
\mathbf{G}_{m\geq m_{1},m_{0}\leq m_{1}}  &  =\frac{e^{ik(m-m_{1}%
)a}\mathcal{T}e^{ik(m_{1}-m_{0})a}}{iv/a}=\mathcal{T}\frac{e^{ik(m-m_{0})a}%
}{iv/a},\\
\mathbf{G}_{m\leq m_{1},m_{0}\geq m_{1}}  &  =\frac{e^{-ik(m-m_{1}%
)a}\mathcal{T}e^{-ik(m_{1}-m_{0})a}}{iv/a}=\mathcal{T}\frac{e^{-ik(m-m_{0})a}%
}{iv/a},\\
\mathbf{G}_{m\leq m_{1},m_{0}\leq m_{1}}  &  =\frac{e^{ika|m-m_{0}|}}%
{iv/a}+\frac{e^{-ik(m-m_{1})a}\mathcal{R}e^{ik(m_{1}-m_{0})a}}{iv/a},\\
\mathbf{G}_{m\geq m_{1},m_{0}\geq m_{1}}  &  =\frac{e^{ika|m-m_{0}|}}%
{iv/a}+\frac{e^{ik(m-m_{1})a}\mathcal{R}e^{-ik(m_{1}-m_{0})a}}{iv/a}.
\end{align}
The above results are also consistent with Eq. (\ref{G_ALL_EXP}).

Finally, when the on-site energies of unit cell $m_{1}$ and unit cell $m_{2}$
($>m_{1}$) are both replaced by $\varepsilon_{0}+\delta$, then each unit cell
becomes a scatterer characterized by the conversion matrix in Eq.
(\ref{CM_EXAMPLE}). The conversion matrix $\mathbb{G}^{(\mathrm{C})}$ of the
composite scatterer $(m_{1}+m_{2})$ is given by Eq. (\ref{G_ALL}). In
particular, the surface elements are obtained from Eq. (\ref{G_BOUNDARY}) as%
\begin{align}
\mathbb{G}_{m_{2}m_{1}}^{(\mathrm{C})}  &  =\mathbb{G}_{m_{1}m_{2}%
}^{(\mathrm{C})}=e^{ik(m_{2}-m_{1})a}\frac{\mathcal{T}(1-e^{2ik(m_{2}-m_{1}%
)a}\mathcal{R}^{2})^{-1}\mathcal{T}}{iv/a},\\
\mathbb{G}_{m_{1}m_{1}}^{(\mathrm{C})}  &  =\mathbb{G}_{m_{2}m_{2}%
}^{(\mathrm{C})}=\mathbb{G}+e^{2ik(m_{2}-m_{1})a}\frac{\mathcal{TR}%
(1-e^{2ik(m_{2}-m_{1})a}\mathcal{R}^{2})^{-1}\mathcal{T}}{iv/a},
\end{align}
from which we can obtain the generalized transmission amplitude across the
composite scatterer as
\begin{equation}
\mathcal{T}^{(\mathrm{C})}\equiv\mathbb{G}_{m_{2},m_{1}}^{(\mathrm{C}%
)}\mathbf{g}^{-1}=\mathcal{T}(1-e^{2ik(m_{2}-m_{1})a}\mathcal{R}^{2}%
)^{-1}e^{ik(m_{2}-m_{1})a}\mathcal{T},
\end{equation}
and the generalized reflection amplitude as%
\begin{equation}
\mathcal{R}^{(\mathrm{C})}\equiv\mathbb{G}_{m_{1}m_{1}}^{(\mathrm{C}%
)}\mathbf{g}^{-1}-1=\mathcal{R}+e^{2ik(m_{2}-m_{1})a}\mathcal{TR}%
(1-e^{2ik(m_{2}-m_{1})a}\mathcal{R}^{2})^{-1}\mathcal{T}.
\end{equation}

\subsection{Chiral tunneling and anomalous focusing in graphene \textit{p-n} junction}

\begin{figure}[t]
\includegraphics[width=\columnwidth,clip]{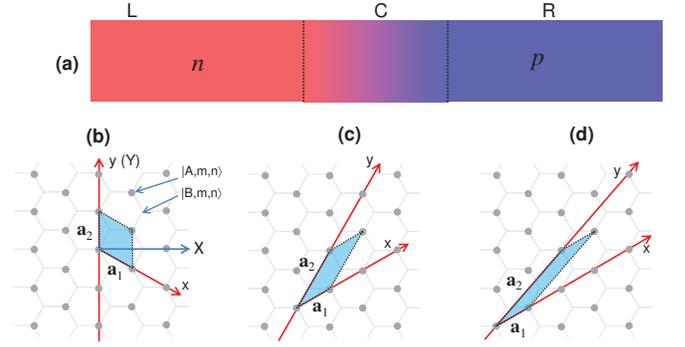}
\caption{(a) Sketch of
graphene \textit{p-n} junction with a smooth interface. Panels (b)-(d) show the choice of
primitive vectors and $x,y$ axes when the interface is along the\ zigzag
direction (b), the armchair direction (c), or a more general direction (d).
The filled dots mark the Bravais lattice ($A$ sublattice) of graphene. The
shaded regions mark the unit cells. In panel (b), the $X$ and $Y$ axes of the
Cartesian coordinate system are also shown.}%
\label{G_PNJ}%
\end{figure}Graphene is a single layer of carbon atoms in a honeycomb lattice
that hosts massless Dirac fermions
\cite{BeenakkerRMP2008,CastroRMP2009,PeresRMP2010,DasSarmaRMP2011}. One of the
unique properties of electrons in graphene is chiral tunneling
\cite{KleinZP1929,KatsnelsonNatPhys2006,CheianovPRB2006}: an electron normally
incident on a potential barrier will always be perfectly transmitted,
independently of its kinetic energy and the height and width of the potential
barrier (for oblique incidence, the transverse momentum serves as a gap-opening mass term, so the transmission is not perfect, and bound states may be created by a 1D potential well \cite{PhysRevB.74.045424,PhysRevA.90.052116}). In recent years, the
chiral tunneling in graphene \textit{p-n} junctions has attracted a lot of attention
(see Ref. \onlinecite{AllainEPJB2011} for a review). Another interesting
phenomenon for electrons in graphene \textit{p-n} junctions is the anomalous focusing
due to negative refraction, which was initially proposed by Veselago for
electromagnetic waves
\cite{VeselagoSPU1968,PendryPRL2000,ZhangNatMater2008,PendryScience2012}: a
spatially diverging pencil of rays is focused to a spatially converging one
during the transition from a medium with positive refractive index across a
sharp interface into a negative index medium. In 2007, Cheianov \textit{et
al}. \cite{CheianovScience2007} proposed that ballistic graphene \textit{p-n}
junctions can also exhibit negative refraction and hence focus the electron
flow:\ in the electron-doped \textit{n} (hole-doped \textit{p}) region, the carrier group
velocity is parallel (anti-parallel) to its momentum, in analogy to light
propagation in a positive (negative) refractive index medium. Ever since then,
there have been a lot of works on the negative refraction in graphene (see
Refs. \onlinecite{ParkNanoLett2008,MoghaddamPRL2010} for examples) and on the
surface of topological insulators \cite{ZhaoPRL2013}. Recently, the Veselago
lens effect in graphene was observed experimentally
\cite{LeeNatPhys2015,ChenScience2016}.

On the theoretical side, most of the previous studies focus on traveling
states and are based on the low-energy continuous model, whose validity is
limited to the vicinity of the Dirac points. A very recent work
\cite{LogemannPRB2015}, based on the tight-binding model, calculates
numerically the propagation of a wave packet in a large but finite graphene
flake to approximate the Klein tunneling and caustics of electron waves. Here
we apply our general GF formalism to study the chiral tunneling in an infinite
graphene \textit{p-n} junction and pay special attention to the \textit{evanescent}
eigenmode. Our approach provides a clear physical picture and allows us to calculate the GF over long distances,
so we further perform both analytical analysis and numerical simulations of dual-probe STM measurements.
Our results demonstrate the possibility of observing the spatially resolved
interference pattern caused by the negative refraction in graphene \textit{p-n}
junctions and further reveal a few interesting features, such as the
distance-independent conductance and its quadratic dependence on the carrier
concentration, as opposed to the linear dependence in uniform graphene.

We consider a graphene \textit{p-n} junction with an interface that can be either sharp
or smooth, as shown in Fig. \ref{G_PNJ}(a). In the tight-binding model, the
interface could align along different crystalline directions. To provide the
simplest description, a \textit{tilted} coordinate system is usually
necessary: we choose one primitive vector $\mathbf{a}_{2}$ (defined as the $y$
axis) of the Bravais lattice of uniform graphene to be parallel to the
interface, and choose the $x$ axis of our tilted coordinate system to be
parallel to the other primitive vector $\mathbf{a}_{1}$; i.e., the two
nonorthogonal unit vectors of the tilted coordinate are $\mathbf{e}_{x}%
\equiv\mathbf{a}_{1}/|\mathbf{a}_{1}|$ and $\mathbf{e}_{y}\equiv\mathbf{a}%
_{2}/|\mathbf{a}_{2}|$, as shown in Figs. \ref{G_PNJ}(b)-\ref{G_PNJ}(d) for the
interface along the zigzag direction, armchair direction, and a more general
direction. This choice of the primitive vectors and the tilted coordinate
system ensures that the lattice Hamiltonian is invariant upon translation by
$|\mathbf{a}_{2}|$ along the $y$ axis, so that the original 2D lattice model
can be reduced to a 1D lattice model.

\subsubsection{Reduction from 2D to 1D}

For specificity, we assume that the interface is along the zigzag direction
[Fig. \ref{G_PNJ}(c)], where $|\mathbf{a}_{1}|=|\mathbf{a}_{2}|=\sqrt
{3}a_{\mathrm{C-C}}\equiv a$  and $a_{\mathrm{C-C}}$ is the C-C bond length. The
vanishingly small spin-orbit coupling in graphene makes the GF
spin-independent, so we neglect the electron spin. In the tight-binding model,
each unit cell of graphene consists of $M=2$ orbitals, i.e., $|A,m,n\rangle$
and $|B,m,n\rangle$ for the unit cell $(m,n)$, where $m$ ($n$) is the index
along the $x$ ($y$) axis of the tilted coordinate and $A,B$ labels the
sublattice [see Fig. \ref{G_PNJ}(c)]. The Hamiltonian $\hat{H}=\hat{H}%
_{0}+\hat{V}$ is the sum of the uniform part
\begin{equation}
\hat{H}_{0}=\sum_{m,n}t(\left\vert A,m+1,n\right\rangle +\left\vert
A,m,n-1\right\rangle +\left\vert A,m,n\right\rangle )\left\langle
B,m,n\right\vert +h.c.)
\end{equation}
and the \textit{p-n} junction potential
\begin{equation}
\hat{V}=\sum_{m,n}V_{m}(\left\vert A,m,n\right\rangle \left\langle
A,m,n\right\vert +\left\vert B,m,n\right\rangle \left\langle B,m,n\right\vert
),
\end{equation}
where $t\approx 3$ eV is the nearest-neighbor hopping constant\cite{CastroRMP2009}. The junction potential $V_{m}$ depends arbitrarily on
$m$ inside the interface ($m_{L}\leq m\leq m_{R}$), but equals $V_{L}$ inside
the left lead $L$ ($m\leq m_{L}-1$) and equals $V_{R}$ inside the right lead
$R$ ($m\geq m_{R}+1$).

Due to the invariance of $\hat{H}$ upon translation by $|\mathbf{a}_{2}|$
along the $y$ axis, the problem can be reduced from 2D to 1D by a Fourier
transform
\begin{equation}
|k_{y}\rangle\equiv\frac{1}{\sqrt{N_{y}}}\sum_{n}e^{ik_{y}{n}a}|n\rangle
,\ \ \ |n\rangle\equiv\frac{1}{\sqrt{N_{y}}}\sum_{k_{y}}e^{-ik_{y}{n}a}%
|k_{y}\rangle, \label{FT}%
\end{equation}
where $N_{y}$ is the number of unit cells along the $y$ axis. In the new
basis, the Hamiltonian $\hat{H}=\sum_{k_{y}}\hat{H}_{\mathrm{1D}}(k_{y}%
)|k_{y}\rangle\langle k_{y}|$ is diagonal with respect to $k_{y}$, where
$\hat{H}_{\mathrm{1D}}(k_{y})$ describes a 1D lattice with the unit cells
labeled by $m$ and each unit cell containing $2$ basis states $|A\rangle
,|B\rangle$. Let us use $\mathbf{R}_{mn}\equiv{m}\mathbf{a}_{1}+n\mathbf{a}%
_{2}$ to denote the Bravais vector of the unit cell $(m,n)$ and use
$\mathbf{G}(\mathbf{R}_{m_{2}n_{2}},\mathbf{R}_{m_{1}n_{1}},E)$ to denote the
retarded GF from the unit cell $(m_{1},n_{1})$ to the unit cell $(m_{2}%
,n_{2})$ of the original 2D system, which is connected to the retarded GF
$\mathbf{G}_{m_{2},m_{1}}(E,k_{y})$ of the 1D lattice from the unit cell
$m_{1}$ to the unit cell $m_{2}$ via a Fourier transform%

\begin{equation}
\mathbf{G}(\mathbf{R}_{m_{2}n_{2}},\mathbf{R}_{m_{1}n_{1}},E)=\frac{1}{N_{y}%
}\sum_{k_{y}}{e^{ik_{y}(n_{2}-n_{1})a}\mathbf{G}_{m_{2},m_{1}}(E,}k_{y}{)},
\label{G2D_FT}%
\end{equation}
where $\mathbf{G}(\mathbf{R}_{m_{2}n_{2}},\mathbf{R}_{m_{1}n_{1}},E)$ and
$\mathbf{G}_{m_{2},m_{1}}(E,k_{y})$ are both $2\times2$ matrices. Below we
consider fixed $E$ and $k_{y}$ and apply our general results to calculate the
GF ${\mathbf{G}_{m_{2},m_{1}}}$ of the 1D lattice, with $E$ and $k_{y}$ omitted
for brevity.

\subsubsection{Green's function of 1D lattice}

In the 1D lattice, the hopping is uniform:
\begin{equation}
\mathbf{H}_{m,m+1}^{(\mathrm{1D})}=(\mathbf{H}_{m+1,m}^{(\mathrm{1D}%
)})^{\dagger}=\mathbf{t}=%
\begin{bmatrix}
0 & 0\\
t & 0
\end{bmatrix}
. \label{T}%
\end{equation}
The unit cell Hamiltonian $\mathbf{H}_{m,m}^{(\mathrm{1D})}=V_{m}%
+\mathbf{h}_{0}$ is the sum of the unit cell Hamiltonian of pristine
graphene,%
\begin{equation}
\mathbf{h}_{0}=%
\begin{bmatrix}
0 & t(1+e^{ik_{y}a})\\
t(1+e^{-ik_{y}a}) & 0
\end{bmatrix}
. \label{H0}%
\end{equation}
and the \textit{p-n} junction potential $V_{m}$. The entire infinite system consists of
a single scatterer ($m_{L}\leq m\leq m_{R}$) connected to two semi-infinite
leads $L$ and $R$ [cf. Fig. \ref{G_PNJ}(a)], whose GFs can be constructed from
the conversion and propagation matrices of the leads and the conversion matrix
$\mathbb{G}$ of the \textit{p-n} interface (see Sec. III\ and Sec. IV).
\begin{figure}[t]
\includegraphics[width=\columnwidth,clip]{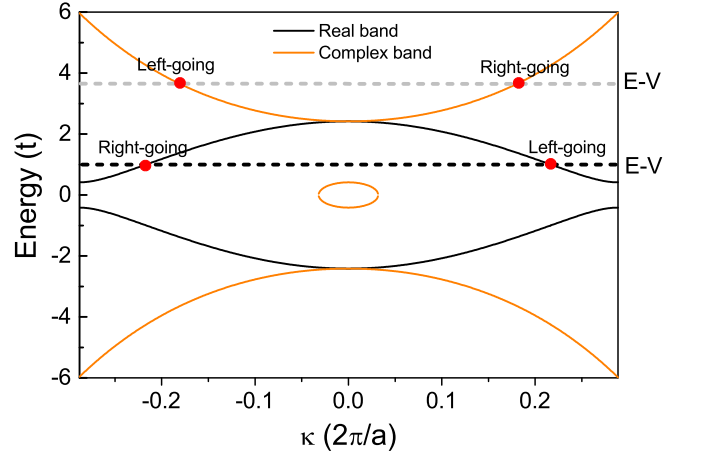}
\caption{Real
energy band (black lines) and complex energy bands (orange lines) of pristine
graphene at a fixed $k_{y}\approx0.14\times2\pi/a$.}%
\label{G_SPECTRA}%
\end{figure}

The remaining issue is to calculate the eigenmodes of each lead, as
characterized by the hopping $\mathbf{t}$ and the unit cell Hamiltonian
$\mathbf{h}=V+\mathbf{h}_{0}$, where $V=V_{L}$ (left lead) or $V_{R}$ (right
lead). Given a complex wave vector $k$, we can solve the eigenvalue problem
Eq. (\ref{MEQ1}) and obtain the energy bands of the lead as $V\pm E_{0}(k)$,
where $E_{0}(k)\equiv t\sqrt{f(k,k_{y})f(-k,-k_{y})}$ and $f(k,k_{y}%
)\equiv1+e^{ik_{y}a}+e^{-ika}$. Here $E_{0}(k)$ is real in two cases: (1)
$k+k_{y}/2=\kappa$; (2)\ $k+k_{y}/2=i\kappa$ or $\pi/a+i\kappa$, where
$\kappa\in\mathbb{R}$. The former gives the real energy bands, while the
latter gives the complex energy bands, as shown in Fig. \ref{G_SPECTRA}.
Conversely, given the energy $E$, we can solve Eq. (\ref{MEQ1}) and obtain a
right-going eigenmode $k_{+},|\Phi_{+}\rangle$ and a left-going eigenmode
$k_{-},|\Phi_{-}\rangle$, where $k_{\pm}$ are the two solutions to
$|E-V|=E_{0}(k)$ or equivalently the two intersection points of $E-V$ with the
real and complex energy bands of pristine graphene. When $E-V$ lies in the
range of the real energy bands (dashed black line in Fig. \ref{G_SPECTRA}),
the wave vectors $k_{\pm}$ are both real and both eigenmodes are traveling
modes. When $E-V$ lies outside the range of the real energy bands (dashed gray
line in Fig. \ref{G_SPECTRA}), the wave vectors $k_{+}=k_{-}^{\ast}$ are
complex and both eigenmodes are evanescent. A more convenient method to obtain
the eigenmodes is to define $\lambda\equiv e^{ika}$ and rewrite Eq.
(\ref{MEQ1}) as
\begin{equation}
\lambda^{2}bt+\lambda\lbrack|b|^{2}+t^{2}-(E-V)^{2}]+tb^{\ast}=0
\end{equation}
with $b\equiv t(1+e^{ik_{y}a})$, so that $\lambda_{\pm}=e^{ik_{\pm}a}$ are
obtained as the two solutions to this quadratic equation for $\lambda$. In
addition to the two \textit{normal} eigenmodes, there are also a pair of
\textit{ideally evanescent} eigenmodes, including a right-going one
$\lambda_{+,0}=e^{ik_{+,0}a}=0$, $|\Phi_{+,0}\rangle=[1,0]^{T}$ and a
left-going one $\lambda_{-,0}=e^{ik_{-,0}a}=\infty$, $|\Phi_{-,0}%
\rangle=[0,1]^{T}$ (see Appendix \ref{APPEND_BULKMODE}). The ideally
evanescent eigenmodes do not propagate, so they do not directly contribute to
the GF, but their existence does influence the generalized transmission and
reflection amplitudes of the normal eigenmodes.

Using the eigenmodes, we can calculate the conversion matrix $\mathbb{G}$ of
the \textit{p-n} interface using Eq. (\ref{GG}) and then obtain the generalized
transmission and reflection amplitudes $\mathcal{S}^{(LL)}$, $\mathcal{S}%
^{(RL)}$, $\mathcal{S}^{(LR)}$, and $\mathcal{S}^{(RR)}$ of the normal
eigenmode from Eqs. (\ref{SRL})-(\ref{SRR}). For $m\neq m_{0}$, the free GF of
the left lead is essentially the sum of the left-going eigenmode $|\Phi
_{-}^{(L)}(m)\rangle$ and the right-going eigenmode $|\Phi_{+}^{(L)}%
(m)\rangle$ [cf. Eq. (\ref{BULK_MODE_DEF}) for their definitions]:%
\begin{equation}
\mathbf{g}_{m,m_{0}}^{(L)}=\frac{a}{iv_{+}^{(L)}}|\Phi_{+}^{(L)}%
(m)\rangle\langle\tilde{\Phi}_{-}^{(L)}|+\frac{a}{-iv_{-}^{(L)}}|\Phi
_{-}^{(L)}(m)\rangle\langle\tilde{\Phi}_{+}^{(L)}|.
\end{equation}
Due to the \textit{p-n} interface, the eigenmode $|\Phi_{+}^{(L)}(m)\rangle$ produces a
reflection wave in the left lead and a transmission wave in the right lead:%
\begin{equation}
|\Psi_{+}^{(L,\mathrm{out})}(m)\rangle=\left\{
\begin{array}
[c]{ll}%
e^{ik_{-}^{(L)}(m-m_{L})a}|\Phi_{-}^{(L)}\rangle\mathcal{S}^{(LL)}%
e^{ik_{+}^{(L)}(m_{L}-m_{0})a} & (m\in L),\\
e^{ik_{+}^{(R)}(m-m_{R})a}|\Phi_{+}^{(R)}\rangle\mathcal{S}^{(RL)}%
e^{ik_{+}^{(L)}(m_{L}-m_{0})a} & (m\in R),
\end{array}
\right.
\end{equation}
so the GF from the left lead to the right lead is essentially the transmission
wave:%
\begin{equation}
\mathbf{G}_{m\in R,m_{0}\in L}=\frac{a}{iv_{+}^{(L)}}|\Psi_{+}%
^{(L,\mathrm{out})}(m)\rangle\langle\tilde{\Phi}_{-}^{(L)}|,
\end{equation}
while the GF inside the left lead is essentially the sum of the incident
eigenmode $|\Phi_{\pm}^{(L)}(m)\rangle$ and the reflection wave:%
\begin{equation}
\mathbf{G}_{m\in L,m_{0}\in L}=\mathbf{g}_{m,m_{0}}^{(L)}+\frac{a}%
{iv_{+}^{(L)}}|\Psi_{+}^{(L,\mathrm{out})}(m)\rangle\langle\tilde{\Phi}%
_{-}^{(L)}|.
\end{equation}

\subsubsection{Green's function of 2D graphene \textit{p-n} junction: anomalous
focusing}
Compared with the standard RGF method, the advantages of our GF method lie in its physical transparency and numerical efficiency. Here we demonstrate the first point by using our method to provide a clear physical picture of the anomalous focusing effect \cite{CheianovScience2007,ParkNanoLett2008,MoghaddamPRL2010,LeeNatPhys2015,ChenScience2016} across the graphene \textit{p-n} junction described by the tight-binding model. For this purpose, we first obtain the GF of the 2D graphene \textit{p-n} junction from the 1D GFs by a Fourier transform [Eq. (\ref{G2D_FT})]. In particular, the GF from the unit
cell $(m_{1},n_{1})$ in the \textit{n} region to the unit cell $(m_{2},n_{2})$ in the \textit{p}
region,
\begin{equation}
\mathbf{G}(\mathbf{R}_{m_{2}n_{2}},\mathbf{R}_{m_{1}n_{1}},E)=\int\frac
{dk_{y}}{2\pi}|\Psi_{+}^{(L,\mathrm{tran})}(\mathbf{R}_{m_{2}n_{2}}%
)\rangle\frac{a}{iv_{+}^{(L)}}\langle\tilde{\Phi}_{-}^{(L)}|, \label{G2D_RL}%
\end{equation}
is essentially the sum of all transmission wave functions%
\begin{equation}
|\Psi_{+}^{(L,\mathrm{tran})}(\mathbf{R}_{m_{2}n_{2}})\rangle\equiv
e^{i\mathbf{k}_{+}^{(R)}\cdot(\mathbf{R}_{m_{2}n_{2}}-\mathbf{R}_{m_{R},0}%
)}|\Phi_{+}^{(R)}\rangle\mathcal{S}^{(RL)}e^{i\mathbf{k}_{+}^{(L)}%
\cdot(\mathbf{R}_{m_{L},0}-\mathbf{R}_{m_{1}n_{1}})}, \label{PHI_OUT}%
\end{equation}
which emanates from the incident eigenmode $|\Phi_{+}^{(L)}(\mathbf{R}%
_{m,n})\rangle=e^{i\mathbf{k}_{+}^{(L)}\cdot(\mathbf{R}_{m,n}-\mathbf{R}%
_{m_{1}n_{1}})}|\Phi_{+}^{(L)}\rangle$ through three steps: propagation to the
left interface $\mathbf{R}_{m_{L},0}$ of the junction with wave vector
$\mathbf{k}_{+}^{(L)}$, transmission across the interface, and propagation
from $\mathbf{R}_{m_{R},0}$ to $\mathbf{R}_{m_{2}n_{2}}$ with wave vector
$\mathbf{k}_{+}^{(R)}$. Here $\mathbf{k}_{\pm}^{(L)}$ ($\mathbf{k}_{\pm}%
^{(R)}$) are the wave vectors of the \textit{normal} eigenmodes in the left
(right) region, i.e., in the tilted coordinate [Fig. \ref{G_PNJ}%
(b)]:$\ \mathbf{k}_{\pm}^{(p)}\cdot\mathbf{e}_{x}=k_{\pm}^{(p)}$ and
$\mathbf{k}_{\pm}^{(p)}\cdot\mathbf{e}_{y}\equiv k_{y}$ ($p=L,R$). The GF from
$\mathbf{R}_{m_{1}n_{1}}$ to $\mathbf{R}_{m_{2}n_{2}}$ can be measured as the
conductance between one STM probe coupled to $\mathbf{R}_{m_{1}n_{1}}$ and
another STM probe coupled to $\mathbf{R}_{m_{2}n_{2}}$ through the
Landauer-B\"{u}ttiker formula \cite{DattaBook1995} ($2e^{2}/h)T(E_{F})$, where
the transmission probability $T(E_{F})\propto|\mathbf{G}(\mathbf{R}%
_{m_{2}n_{2}},\mathbf{R}_{m_{1}n_{1}},E_{F})|^{2}$. Therefore, Eq.
(\ref{G2D_RL}) directly connects the transmission wave function to the
experimentally measurable conductance and hence provides a clear physical
picture for observing the anomalous focusing in dual-probe STM measurements,
and further reveals some interesting effects.

Let us consider a sharp, symmetric interface at $m_{L}=m_{R}=0$ and
$V_{R}=-V_{L}=V_{0}>0$. In this case, the transmission wave simplifies to%
\begin{equation}
|\Psi_{+}^{(L,\mathrm{tran})}(\mathbf{R}_{m_{2}n_{2}})\rangle\equiv
e^{i(\mathbf{k}_{+}^{(R)}\cdot\mathbf{R}_{m_{2}n_{2}}-\mathbf{k}_{+}%
^{(L)}\cdot\mathbf{R}_{m_{1}n_{1}})}|\Phi_{+}^{(R)}\rangle\mathcal{S}^{(RL)}.
\label{PHI_OUT_2}%
\end{equation}
Here $\mathbf{k}_{+}^{(R)}$, $\mathbf{k}_{+}^{(L)}$, $|\Phi_{+}^{(R)}\rangle$,
and $\mathcal{S}^{(RL)}$ all depend on $k_{y}$ weakly. The strongest
dependence on $k_{y}$ comes from the phase factor $e^{i(\mathbf{k}_{+}%
^{(R)}\cdot\mathbf{R}_{m_{2}n_{2}}-\mathbf{k}_{+}^{(L)}\cdot\mathbf{R}%
_{m_{1}n_{1}})}$, which usually oscillates rapidly as a function of $k_{y}$
when $\mathbf{R}_{m_{2}n_{2}}$ and $\mathbf{R}_{m_{1}n_{1}}$ are far away.
However, when the energy of the incident electron lies midway in between the
Dirac point of the \textit{n}\ region and the Dirac point of the \textit{p} region (i.e.,
$E=E_{F}=0$), in the \textit{Cartesian} coordinate system spanned by the orthogonal unit vectors $\mathbf{e}_{X}$ and $\mathbf{e}_{Y}$ [see Fig. \ref{G_PNJ}(b)], the
electron-hole symmetry of graphene dictates that the Fermi wave vector
$\mathbf{k}_{+}^{(L)}$ of the right-going eigenmode in the \textit{n} region and the
Fermi wave vector $\mathbf{k}_{+}^{(R)}$ of the right-going eigenmode in the \textit{p}
region to have the same component along the \textit{p-n} interface (i.e.,
$\mathbf{k}_{+}^{(L)}\cdot\mathbf{e}_{Y}=\mathbf{k}_{+}^{(R)}\cdot
\mathbf{e}_{Y}$), but opposite components perpendicular to the \textit{p-n} interface
(i.e., $\mathbf{k}_{+}^{(L)}\cdot\mathbf{e}_{X}=-\mathbf{k}_{+}^{(R)}%
\cdot\mathbf{e}_{X}$). Therefore, when $\mathbf{R}_{m_{1}n_{1}}$ and
$\mathbf{R}_{m_{2}n_{2}}$ are mirror symmetric about the \textit{p-n} interface, the
rapidly oscillating phase factor equals unity for all $k_{y}$. In this case,
all the transmitted waves have nearly the same phase factor for all $k_{y}$, so
they contribute constructively to the GF. This corresponds to electrons
flowing out of an electron source at $\mathbf{R}_{m_{1}n_{1}}$ being focused
to $\mathbf{R}_{m_{2}n_{2}}$, i.e., the anomalous focusing
\cite{CheianovScience2007}. According to the Landauer-B\"{u}ttiker formula,
the constructive enhancement of the GF could be detected as an enhanced
conductance in dual-probe STM measurements.

In addition to locally enhancing the GF, the constructive interference of all
the transmission waves also gives rise to two interesting behaviors. First,
the phase factor $e^{i(\mathbf{k}_{+}^{(R)}\cdot\mathbf{R}_{m_{2}n_{2}%
}-\mathbf{k}_{+}^{(L)}\cdot\mathbf{R}_{m_{1}n_{1}})}$ and hence the
transmission wave and the GF remain invariant when $\mathbf{R}_{m_{2}n_{2}}$
and $\mathbf{R}_{m_{1}n_{1}}$ are moved equally but in opposite directions
perpendicular to the \textit{p-n} interface. This would give rise to
distance-independent conductance. Second, since each transmission wave
contributes constructively to the GF, we have $\mathbf{G}(\mathbf{R}%
_{m_{2}n_{2}},\mathbf{R}_{m_{1}n_{1}},E_{F})\propto$ density of states on the
Fermi surface $\propto V_{0}\propto$ carrier concentration. Thus the locally
enhanced conductance should increase quadratically with increasing doping
level $V_{0}$ (or equivalently the carrier concentration), in contrast to the
linear dependence in uniform graphene \cite{SettnesPRL2014}. These points will be verified in our subsequent numerical simulations of the dual-probe STM measurements, which provide a useful tool, with high spatial resolution, to measure such local transport properties and detect possible zero-energy bound states of the Dirac fermions caused by suitable 2D potential well \cite{PhysRevLett.102.226803,PhysRevB.92.165401}.
\subsubsection{Comparison of the standard RGF and our GF approach}
Compared with the standard RGF that treats the \textit{entire} central region numerically [see Fig. \ref{G_SETUP}(a) for an example], an important advantage of our approach is that it fully utilizes the translational invariance of all the periodic subregions [even if they lie inside the central region, such as the middle lead in Fig. \ref{G_SETUP}(a)] to treat these subregions semianalytically, so that only the \textit{truly disordered} subregions need numerical treatment. Therefore, our GF approach is more efficient if the central region contains periodic subregions; otherwise the two approaches are equally efficient. Here we calculate $G(\mathbf{R}_{2},\mathbf{R}_{1},E)$ across a sharp graphene \textit{p-n} junction using these two methods to demonstrate their equivalence and highlight the numerical efficiency of our approach. In the calculation, ${\bf R}_1$ is fixed at a randomly chosen A-sublattice site in the \textit{n} region; ${\bf R}_2$ is swept over all the B-sublattice sites along the $x$ axis from the \textit{n} region to the \textit{p} region. The range of the sweep is from $0.25 \mu$m on the left of ${\bf R}_1$ to $0.75 \mu$m on the right of ${\bf R}_1$. For the RGF method, the central region (infinite along the \textit{p-n} interface) is the smallest region that encloses ${\bf R}_1$, ${\bf R}_2$, and the \textit{p-n} interface. For our GF method, the scatterer region (infinite along the \textit{p-n} interface) consists of one slice at the \textit{p-n} interface. The numerical results from the two approaches always agree with each other up to the machine accuracy. The time cost, however, differs by two orders of magnitude: with 90 Intel cores, the time cost of the standard RGF approach varies from 260 s to 540 s, depending on the position of ${\bf R}_1$ relative to the \textit{p-n} interface, while the time cost of our approach is always less than 3 s. Similar speedup is expected in long multilayer structures with sharp interfaces, such as quantum wells, superlattices, or sharp \textit{p-n-p} junctions.
\subsubsection{Numerical examples}

\begin{figure}[t]
\includegraphics[width=\columnwidth]{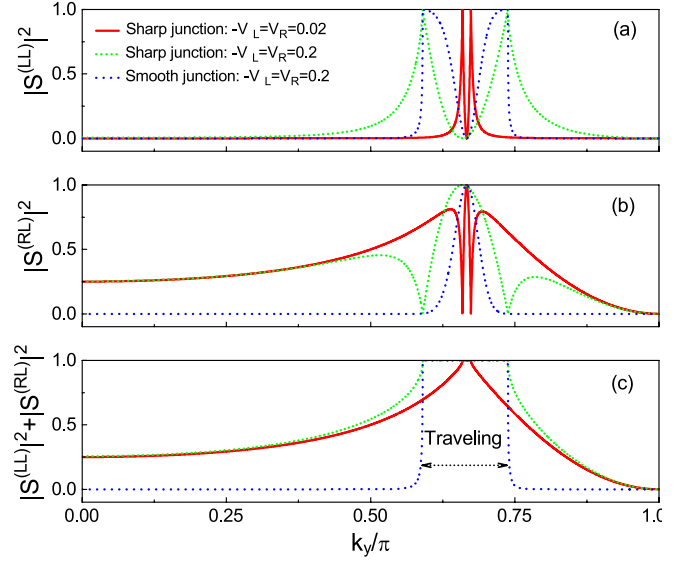}
\caption{Transmission and
reflection of a right-going eigenmode of energy $E=0$ incident from the \textit{n}
region of a sharp (solid lines and dotted lines) or 5nm wide smooth (dashed
lines), symmetric (i.e., $-V_{L}=V_{R}=V_{0}$) graphene \textit{p-n} junction. For
$V_{0}=0.2$, the range of $k_{y}$ in which the eigenmode is traveling is
marked by the double arrow. For $V_{0}=0.02$, the range of $k_{y}$ in which
the eigenmode is traveling is much narrower.}%
\label{G_TRANS}%
\end{figure}

In the following (main text and all the figures), we always use the C-C bond length of graphene
$a_{\mathrm{C-C}}=0.142$ nm as the unit of length and the nearest-neighbor hopping
amplitude $t=3$ eV as the unit of energy. For specificity, we focus on
symmetric graphene \textit{p-n} junctions with $V_{R}=-V_{L}=V_{0}>0$. Unless
explicitly specified, we always take a typical doping level $V_{0}=0.2$ and
set the energy $E=E_{F}=0$, so we denote $G(\mathbf{R}_{2},\mathbf{R}_{1},E)$
by $G(\mathbf{R}_{2},\mathbf{R}_{1})$ for brevity.

As shown in Fig. \ref{G_TRANS}, at $V_{0}=0.2$, the tunneling of a right-going
traveling eigenmode reproduces the well-known results from the continuum model
\cite{AllainEPJB2011}, such as the perfect transmission at normal incidence.
For the evanescent eigenmode in a sharp \textit{p-n} junction, however, $|\mathcal{S}%
^{(RL)}|^{2}$ shows a peak, indicating enhanced tunneling of certain
evanescent states. When the Fermi level is tuned closer to the Dirac point,
i.e., for $V_{0}=0.02$, this enhanced tunneling becomes more pronounced and
may be observed by dual-probe STM measurements.

\begin{figure*}[t]
\includegraphics[width={1.0\textwidth} ]{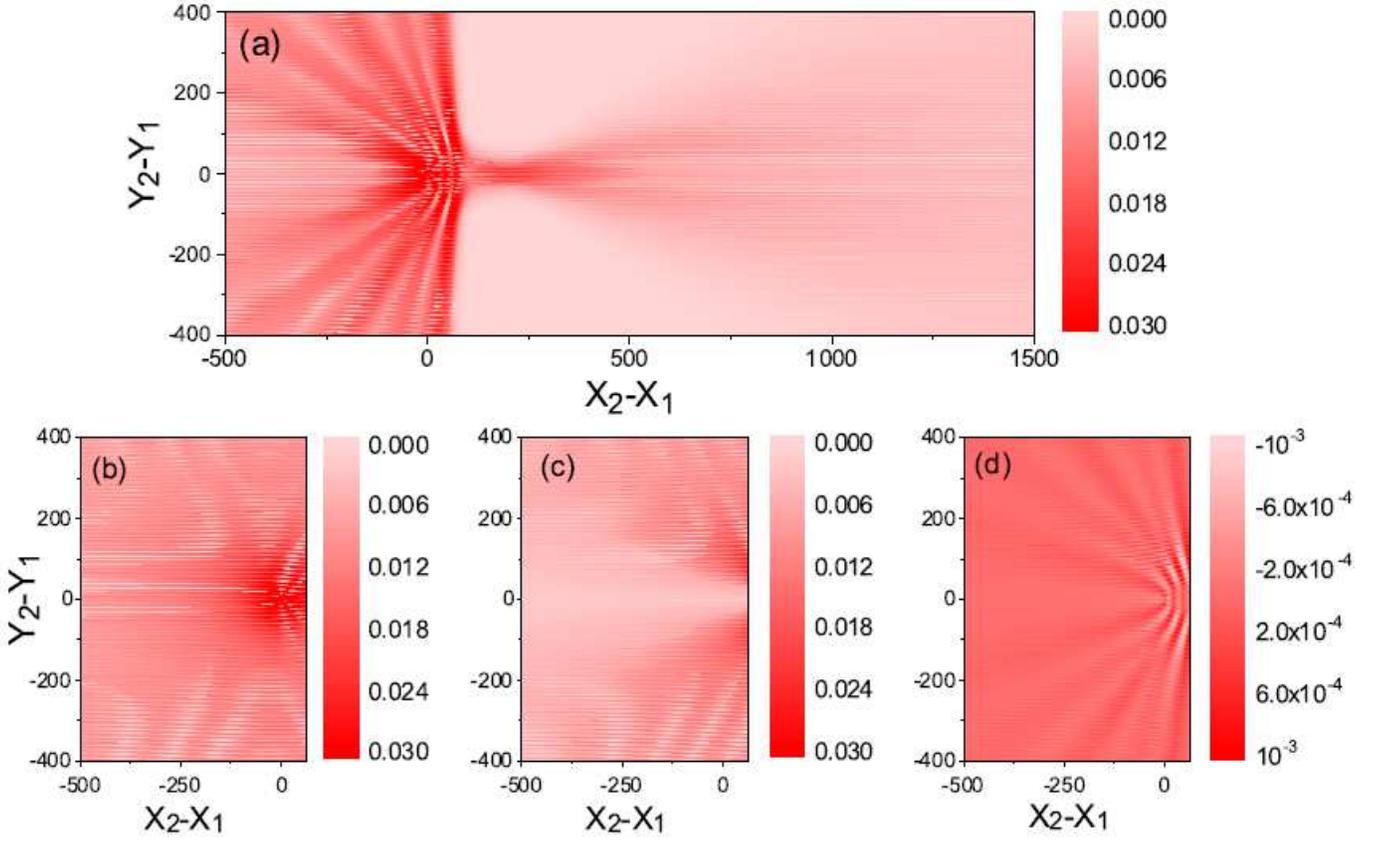}
\caption{Contributions of different scattering channels to the Green's
function $G(\mathbf{R}_{2},\mathbf{R}_{1})$ for a smooth linear graphene \textit{p-n}
junction extending from $65$ to $135$ along the $x$ axis (junction width
$\approx10$ nm). Here $\mathbf{R}_{1}\equiv(X_{1},Y_{1})$ is chosen to be
fixed at $(0,0)$ on the $A$ sublattice, while $\mathbf{R}_{2}\equiv
(X_{2},Y_{2})$ is swept over all the $B$ sublattice sites of the entire
structure. (a) Contour plot of $|G(\mathbf{R}_{2},\mathbf{R}_{1})|$ vs
$\mathbf{R}_{2}-\mathbf{R}_{1}$. Panels (b) and (c) extract the contributions to
$|G(\mathbf{R}_{2},\mathbf{R}_{1})|$ from the incident wave and reflection
wave, respectively. Panel (d) shows the contribution to $|G(\mathbf{R}%
_{2},\mathbf{R}_{1})|^{2}$ due to the interference between the incident wave
and the reflection wave. }%
\label{GFtu12}%
\end{figure*}

Next we visualize the contributions from different scattering channels to the
GF and their interference. For a smooth junction, the spatial map of the GF
[Fig. \ref{GFtu12}(a)] shows imperfect focusing \cite{CheianovScience2007} due
to negative refraction across a finite-width \textit{p-n} junction.
Let us consider
$\mathbf{R}_{m_{2}n_{2}}$ and $\mathbf{R}_{m_{1}n_{1}}$ both in the \textit{n} region
and $\mathbf{R}_{m_{2}n_{2}}$ on the right of $\mathbf{R}_{m_{1}n_{1}}$ (i.e.,
$m_{2}>m_{1}$); the GF
\begin{align}
\mathbf{G}(\mathbf{R}_{m_{2}n_{2}},\mathbf{R}_{m_{1}n_{1}})  & =\int
\frac{dk_{y}}{2\pi}\frac{a}{iv_{+}^{(L)}}\nonumber\\
& \times\lbrack|\Phi_{+}^{(L)}(\mathbf{R}_{m_{2}n_{2}})\rangle+|\Psi
_{+}^{(L,\mathrm{refl})}(\mathbf{R}_{m_{2}n_{2}})\rangle]\langle\tilde{\Phi}_{-}%
^{(L)}|
\end{align}
is essential the sum of all right-going eigenmodes $|\Phi_{+}^{(L)}%
(\mathbf{R}_{m_{2}n_{2}})\rangle$ and all reflection waves
\begin{equation}
|\Psi_{+}^{(L,\mathrm{refl})}(\mathbf{R}_{m_{2}n_{2}})\rangle\equiv
e^{i\mathbf{k}_{-}^{(L)}\cdot(\mathbf{R}_{m_{2}n_{2}}-\mathbf{R}_{m_{L},0}%
)}|\Phi_{-}^{(L)}\rangle\mathcal{S}^{(LL)}e^{i\mathbf{k}_{+}^{(L)}%
\cdot(\mathbf{R}_{m_{L,0}}-\mathbf{R}_{m_{1}n_{1}})}.
\end{equation}
The incident wave contribution coincides with that of pristine graphene [Fig. \ref{GFtu12}(b)]. The reflection wave
contribution [Fig. \ref{GFtu12}(c)] tends to vanish perpendicularly to the
junction interface, indicative of the chiral tunneling. The interference
between the incident and reflection waves [Fig. \ref{GFtu12}(d)] is
responsible for the interference pattern in the total GF [Fig. \ref{GFtu12}%
(a)], which would be directly manifested in dual-probe STM measurements.

\begin{figure}[t]
\includegraphics[width=\columnwidth, clip]{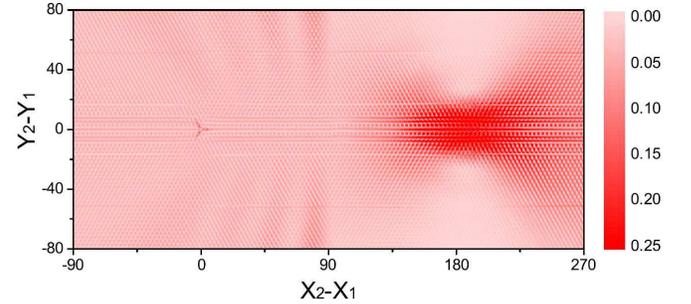}
\caption{Spatial map of the scaled conductance (i.e., the transmission coefficient $T_{12}$ times $|\mathbf{R}_{2}-\mathbf{R}_{1}|$) in a sharp graphene \textit{p-n} junction as a function of $\mathbf{R}_{2}-\mathbf{R}_{1}$. Here $\mathbf{R}_{1}\equiv(X_{1},Y_{1})$ is fixed at the $A$ sublattice of the unit
cell $(0,0)$ and $\mathbf{R}_{2}\equiv(X_{2},Y_{2})$ is swept over all the lattice sites (including both $A$ and $B$ sublattices) of the entire structure.}%
\label{condmap}%
\end{figure}

Finally we simulate dual-probe STM measurements
\cite{SettnesPRL2014,SettnesPRB2014} over the graphene \textit{p-n} junction at zero
temperature.   Following Refs. \onlinecite{SettnesPRL2014,SettnesPRB2014}, we
assume that each probe couples to a single carbon site for simplicity, so the
Landauer-B\"{u}ttiker formula \cite{DattaBook1995} gives the interprobe
conductance as $\sigma(\mathbf{R}_{2},\mathbf{R}_{1})=(2e^{2}/h)T_{12}%
(E_{F})=\Gamma_{1}\Gamma_{2}|\mathbf{\bar{G}(R}_{2},\mathbf{R}_{1},E_{F})|^{2}$, where
$\mathbf{\bar{G}(R}_{2},\mathbf{R}_{1},E_{F})$ is the GF incorporating the
self-energy corrections from the STM probes and $\Gamma_{1,2}$ are coupling constants between the STM probes and the graphene. Usually, $\Gamma_{1,2}$ have a sensitive exponential dependence on the distance between the STM probe and the graphene sample, but their specific values do not affect the shape of the signal. Therefore, following Ref. \onlinecite{SettnesPRB2014}, we always rescale the maximum of $T_{12}$ to unity. In Fig. \ref{condmap}, the real-space conductance map shows a pronounced focusing due to negative
refraction \cite{CheianovScience2007}. Other observable electron optics
features include the high transparency of the junction near normal incidence,
i.e., chiral tunneling \cite{KatsnelsonNatPhys2006}, and the interference
pattern between the incident and reflection waves. Recently, negative
refraction in graphene \textit{p-n} junctions was observed \cite{LeeNatPhys2015}, but
the measurement via macroscopic contacts only gives a spatially averaged
result. Here our simulation shows that dual-probe STM measurements can further
provide spatially resolved interference pattern; i.e., dual-probe STM could be
an ideal experimental technique for studying local transport and quantum
interference phenomenon. \begin{figure}[t]
\includegraphics[width=\columnwidth]{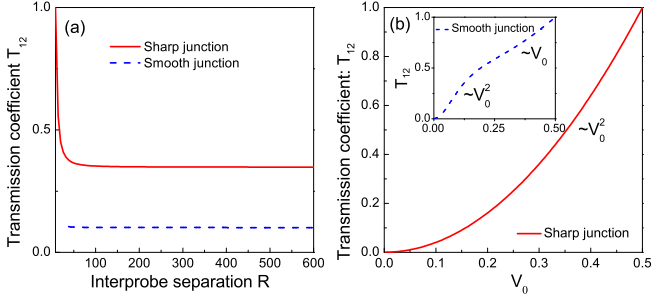}
\caption{Transmission coefficient $T_{12}$ between two STM probes at mirror
symmetric locations about the junction interface. (a) $T_{12}$ vs. interprobe
distance. (b) $T_{12}$ vs. $V_{0}$. Here, the width of the smooth junction
is 5 nm for (a) and inset of (b), and the maximum of $T_{12}$ is always rescaled to 1 in each panel.}%
\label{depend}%
\end{figure}

Now we demonstrate numerically some interesting features of the dual-probe STM
measurements in graphene \textit{p-n} junctions [see the physical analysis following
Eq. (\ref{PHI_OUT_2})]. First, when the two probes are mirror symmetric about
the junction interface, $T_{12}$ is nearly independent of the interprobe
distance [Fig. \ref{depend}(a)], in contrast to the $1/R$ decay in uniform
graphene \cite{SettnesPRL2014}. For a sharp junction, this behavior can be
attributed to the anomalous focusing \cite{CheianovScience2007}. However, the
existence of the same behavior for a smooth junction interface indicates that
it has a different physical origin, i.e., the cancellation of the propagation
phases due to the matching of the electron Fermi surfaces in the \textit{n} region and
the hole Fermi surface in the \textit{p} region \cite{ZhangPRB2016}. The
distance-independent response could change qualitatively the spatial scaling
of many physical quantities, such as the Friedel oscillation induced by an
impurity and the carrier-mediated RKKY interaction between two localized
magnetic moments. Second, the conductance across a sharp junction scales
quadratically with the junction potential:$\ T_{12}\propto V_{0}^{2}$ [red
line in Fig. \ref{depend}(b)], which differs qualitatively from the linear
scaling $T_{12}\propto V_{0}$ of uniform graphene \cite{SettnesPRL2014}. When
the junction becomes smooth [inset of Fig. \ref{depend}(b)], the quadratic
dependence still persists for small $V_{0}$ (electron's Fermi wavelength $\gg$
junction width), but recovers the linear scaling behavior in uniform graphene
for large $V_{0}$. This can be attributed to the gradual destruction of the
anomalous focusing when the junction width becomes larger than the Fermi
wavelength \cite{CheianovPRB2006}.

\section{Conclusions}

We have presented a numerically efficient and physically transparent formalism
to calculate and understand the Green's function (GF) of a general layered
structure. In contrast to the commonly used recursive GF method that directly
calculates the GF through the Dyson equations, our approach converts the
calculation of the GF to the generation and subsequent propagation of a
scattering wave function emanating from a local excitation. This viewpoint
provides analytical expressions of the GF in terms of a generalized scattering
matrix. This identifies the contributions of individual scattering channels to
the GF and hence allows this information to be extracted quantitatively from
dual-probe STM experiments. We further derive an analytical rule to construct
the GF of a general layered system, which could significantly reduce the
computational time cost and enable quantum transport calculations for large
samples. Application of this formalism to simulate the two-dimensional
conductance map of a realistic graphene \textit{p-n} junction demonstrates the
possibility of observing the spatial interference caused by negative
refraction and further reveals a few interesting features, such as the
distance-independent conductance and its quadratic dependence on the carrier
concentration, as opposed to the linear dependence in uniform graphene.
In addition to conventional mesoscopic quantum transport, it would be interesting to apply our GF approach to the investigation other electron interference phenomena, such as the carrier-mediated RKKY interaction between local magnetic moments, the impurity-induced Friedel oscillation, and using dual-probe STM measurements to detect possible zero-energy bound states in graphene caused by suitable 2D potentials \cite{PhysRevLett.102.226803,PhysRevB.92.165401}.

\section{ACKNOWLEDGMENTS}

This work was supported by the MOST of China (Grants
No. 2015CB921503 and No. 2014CB848700), the NSFC
(Grants No. 11274036, No. 11322542, No. 11434010, and No.11504018), and the NSFC program for "Scientific Research Center" (Grant No. U1530401). We acknowledge the computational support from the Beijing Computational Science
Research Center (CSRC).

\appendix

\section{Calculation of eigenmodes}

\label{APPEND_BULKMODE}

When the determinant of the $M\times M$ \ hopping matrix $\mathbf{t}^{\dagger
}$ or $\mathbf{t}$ is nonzero, the $2M$ eigenmodes can be obtained by letting
$\lambda\equiv e^{ika}$ and rewriting Eq. (\ref{MEQ1}) or equivalently
$[-\mathbf{t}^{\dagger}+\lambda(z-\mathbf{h})-\lambda^{2}\mathbf{t}%
]|\Phi\rangle\equiv0\mathbf{\ }$as a generalized $2M\times2M$ eigenvalue
problem%
\begin{equation}%
\begin{bmatrix}
0 & 1\\
-\mathbf{t}^{\dagger} & z-\mathbf{h}%
\end{bmatrix}%
\begin{bmatrix}
|\Phi\rangle\\
\lambda|\Phi\rangle
\end{bmatrix}
=\lambda%
\begin{bmatrix}
1 & 0\\
0 & \mathbf{t}%
\end{bmatrix}%
\begin{bmatrix}
|\Phi\rangle\\
\lambda|\Phi\rangle
\end{bmatrix}
, \label{GEIG}%
\end{equation}
where $z=E+i0^{+}$. In the numerical calculation, we remove the infinitesimal
imaginary part of $z$, i.e., $z=E$. Then Eq. (\ref{GEIG}) is solved for $2M$
solutions $\{\lambda\equiv e^{ika},|\Phi\rangle\}$, where $k$ could be either
real (traveling modes)\ or complex (evanescent modes). Next the $2M$
eigenmodes are classified into $M$ right-going ones and $M$ left-going ones:
the former consists of traveling modes (i.e., $\operatorname{Im}k=0$) with a
positive group velocity and evanescent modes decaying exponentially to the
right (i.e., $\operatorname{Im}k>0$), while the latter consists of traveling
modes (i.e., $\operatorname{Im}k=0$) with a negative group velocity and
evanescent modes decaying exponentially to the left (i.e., $\operatorname{Im}%
k<0$). There is an alternative but less accurate method to calculate and
classify the eigenmodes. In this approach, the infinitesimal imaginary part of
$z$ is replaced by a finite but small positive number $\eta$, i.e.,
$z=E+i\eta$. Then Eq. (\ref{GEIG}) is solved for the $2M$ solutions
$\{\lambda\equiv e^{ika},|\Phi\rangle\}$. Next, according to the sign of
$\operatorname{Im}k$, the $2M$ eigenmodes are classified into $M$ right-going
ones with $\operatorname{Im}k>0$ and $M$ left-going ones with
$\operatorname{Im}k<0$. In the limit $\eta\rightarrow0^{+}$, the imaginary
part $\operatorname{Im}k$ may either vanish (i.e., traveling modes) or remain
finite (i.e., evanescent modes). Obviously, the second approach is accurate
only in the limit of $\eta\rightarrow0^{+}$, so we always use the first
approach in the main text.

When the determinant of the hopping matrix $\mathbf{t}$ vanishes, solving Eq.
(\ref{GEIG}) would give some trivial evanescent eigenmodes. Suppose that
$M_{0}$ out of the $M$ eigenvalues of the hopping matrix $\mathbf{t}^{\dagger
}$ or $\mathbf{t}$ are zero, and the corresponding eigenvectors are
$\{|\Phi_{+,\alpha}\rangle\}$ (for $\mathbf{t}^{\dagger}$) and $\{|\Phi
_{-,\alpha}\rangle\}$ (for $\mathbf{t}$), where $\alpha=1,2,\cdots,M_{0}$.
Then there would be $2M_{0}$ trivial evanescent eigenmodes, including $M$
right-going ones $\lambda_{+,\alpha}=0,|\Phi_{+,\alpha}\rangle$ and $M_{0}$
left-going ones $\lambda_{-,\alpha}=\infty$, $|\Phi_{-,\alpha}\rangle$. The
former correspond to $|\Phi(m-1)\rangle=|\Phi_{+,\alpha}\rangle,$
$|\Phi(m)\rangle=|\Phi(m+1)\rangle=0$ in Eq. (\ref{SE}), while the latter
correspond to $|\Phi(m+1)\rangle=|\Phi_{-,\alpha}\rangle,$ $|\Phi
(m)\rangle=|\Phi(m-1)\rangle=0$ in Eq. (\ref{SE}). As a result, propagation
matrices $\mathbf{P}_{-}$ and $\mathbf{P}_{+}^{-1}$ both diverge. However,
this does not affect our formulas, which only contain $\mathbf{P}_{+}$ and
$\mathbf{P}_{-}^{-1}$ due to causality. The only problem is that for a trivial
evanescent eigenmode $(s,\alpha)$, the generalized group velocity
$v_{s,\alpha}$ [Eq. (\ref{VSA})] is not well defined. For example, when
$\lambda_{+,\alpha}=0$ and hence $\lambda_{-,\alpha}=\infty$, the generalized
velocity $v_{+,\alpha}=v_{-,\alpha}^{\ast}=-ia\lambda_{-,\alpha}^{\ast}%
\langle\Phi_{-,\alpha}|\mathbf{t}^{\dagger}|\Phi_{+,\alpha}\rangle$ involves
the product of $\lambda_{-,\alpha}^{\ast}=\infty$ and $\mathbf{t}^{\dagger
}|\Phi_{+,\alpha}\rangle=0$. This singular problem can be avoided by adding
sufficiently small numbers $\{\epsilon\}$ to $\mathbf{t}$ and $\mathbf{t}%
^{\dagger}$ to make their determinant nonzero and take the limit
$\{\epsilon\rightarrow0\}$ at the end of the calculation.

Taking the graphene junction as an example, the 1D left or right lead is
characterized by the hopping matrix $\mathbf{t}$ and unit cell Hamiltonian
$V+\mathbf{h}_{0}$, where $V=V_{L}$ (left lead) or $V_{R}$ (right lead) and
$\mathbf{h}_{0}$, $\mathbf{t}$ are given by Eqs.\ (\ref{T}) and (\ref{H0}).
Here the hopping matrix $\mathbf{t}$ has one zero eigenvalue with eigenvector
$|\Phi_{-,0}\rangle=[0,1]^{T}$, while $\mathbf{t}^{\dagger}$ has one zero
eigenvalue with eigenvector $|\Phi_{+,0}\rangle=[1,0]^{T}$. This gives rise to
two trivial evanescent eigenmodes: the right-going one $\lambda_{+,0}%
=0,|\Phi_{+,0}\rangle$ and the left-going one $\lambda_{-,0}=\infty$,
$|\Phi_{-,0}\rangle$, for which the generalized group velocities
$v_{+,0}=v_{-,0}^{\ast}$ are not well defined. To cure this problem, we add a
small number $\epsilon$ to the off-diagonal of the hopping matrix, so that
$\mathbf{t}$ and $\mathbf{t}^{\dagger}$ become%

\begin{equation}
\mathbf{t}(\epsilon)=%
\begin{bmatrix}
0 & \epsilon\\
t & 0
\end{bmatrix}
,\ \mathbf{t}^{\dagger}(\epsilon)=%
\begin{bmatrix}
0 & t\\
\epsilon & 0
\end{bmatrix}
.
\end{equation}
Then using the first-order perturbation theory, we obtain
\begin{align}
\lambda_{+,0}(\epsilon)  &  \approx\frac{-\epsilon}{t\left(  e^{-iak_{y}%
}+1\right)  },\ |\Phi_{+,0}(\epsilon)\rangle\approx%
\begin{bmatrix}
1\\
\frac{-\epsilon E}{t^{2}(e^{-iak_{y}}+1)}%
\end{bmatrix}
,\\
\lambda_{-,0}(\epsilon)  &  \approx\frac{t\left(  e^{iak_{y}}+1\right)
}{-\epsilon},\ \ |\Phi_{-,0}(\epsilon)\rangle\approx%
\begin{bmatrix}
\frac{-\epsilon E}{t^{2}(e^{iak_{y}}+1)}\\
1
\end{bmatrix}
.
\end{align}
Substituting into Eq. (\ref{VSA}) and taking the limit $\epsilon\rightarrow0$
gives
\begin{equation}
v_{+,0}=v_{-,0}^{\ast}=iat(e^{-iak_{y}}+1).
\end{equation}

\section{Scattering of a partial wave}

\label{APPEND_EQUIVALENCE}

Let us consider a scatterer connected to a semi-infinite left lead (whose
conversion matrix is $\mathbf{g}$) and prove that a right-going incident
partial wave $|\Phi_{\mathrm{in}}(m)\rangle$ is equivalent to a local
excitation $|\Phi_{\mathrm{loc}}\rangle_{m_{0}}\equiv\mathbf{g}^{-1}%
|\Phi_{\mathrm{in}}(m_{0})\rangle$ at an arbitrary site $m_{0}\leq m_{L}$
($m_{L}$ is the left surface of the scatterer), in the sense that they produce
the same scattering state at $m\geq m_{0}$. For this purpose, we use
$|\Psi(m)\rangle$ for the conventional scattering state emanating from the
incident wave $|\Phi_{\mathrm{in}}(m)\rangle$ and $|\Phi(m)\rangle$ for the
scattering state emanating from the local excitation $|\Phi_{\mathrm{loc}%
}\rangle_{m_{0}}$ at $m_{0}$. We notice that $|\Psi(m)\rangle$ and
$|\Phi(m)\rangle$ obey the same Schr\"{o}dinger equation for $m\geq m_{0}+1$,
and the same uniform Schr\"{o}dinger equation
\begin{equation}
-\mathbf{t}^{\dagger}|\Phi(m-1)\rangle+(z-\mathbf{h})|\Phi(m)\rangle
-\mathbf{t}|\Phi(m+1)\rangle=0
\end{equation}
for $m\leq m_{0}-1$. The difference lies at $m=m_{0}$:
\begin{align}
&  -\mathbf{t}^{\dagger}|\Psi(m_{0}-1)\rangle+(z-\mathbf{H}_{m_{0},m_{0}%
})|\Psi(m_{0})\rangle-\mathbf{H}_{m_{0},m_{0}+1}|\Psi(m_{0}+1)\rangle
\nonumber\\
&  =0,\\
&  -\mathbf{t}^{\dagger}|\Phi(m_{0}-1)\rangle+(z-\mathbf{H}_{m_{0},m_{0}%
})|\Phi(m_{0})\rangle-\mathbf{H}_{m_{0},m_{0}+1}|\Phi(m_{0}+1)\rangle
\nonumber\\
&  =|\Phi_{\mathrm{loc}}\rangle_{m_{0}},
\end{align}
and the boundary conditions: $|\Phi(m)\rangle$ should be finite at
$m\rightarrow-\infty$, while the right-going part of $|\Psi(m)\rangle$ should
equal the incident wave on the left of the scatterer, i.e., $|\Psi
_{+}(m)\rangle=|\Phi_{\mathrm{in}}(m)\rangle$ for $m\leq m_{L}$. The former
gives $|\Phi(m_{0}-1)\rangle=\mathbf{P}_{-}^{-1}|\Phi(m_{0})\rangle$, while
the latter gives
\begin{align}
|\Psi(m_{0}-1)\rangle &  =\mathbf{P}_{+}^{-1}|\Psi_{+}(m_{0})\rangle
+\mathbf{P}_{-}^{-1}|\Psi_{-}(m_{0})\rangle\\
&  =\mathbf{P}_{-}^{-1}|\Psi(m_{0})\rangle+(\mathbf{P}_{+}^{-1}-\mathbf{P}%
_{-}^{-1})|\Phi_{\mathrm{in}}(m_{0})\rangle.
\end{align}
Using these relations to eliminate $|\Phi(m_{0}-1)\rangle$ and $|\Psi
(m_{0}-1)\rangle$ from the equations for $|\Phi(m)\rangle$ and $|\Psi
(m)\rangle$ at $m=m_{0}$, and using Eq. (\ref{EQUALITY}), we see that they
become identical and contains the same source $|\Phi_{\mathrm{loc}}%
\rangle_{m_{0}}$. Therefore, $|\Phi(m)\rangle|_{m\geq m_{0}}$ and
$|\Psi(m)\rangle|_{m\geq m_{0}}$ obeys exactly the same set of closed
equations and natural boundary conditions (i.e., they should be finite for
$m\rightarrow+\infty$); thus they are identical. Applying this equivalence
principle to a scatterer connected to a semi-infinite left lead $L$ and a
semi-infinite right lead $R$ gives Eqs. (\ref{PHI_S1}) and (\ref{PHI_S2}) of
the main text.

\section{Green's function of one scatterer: simple examples}

\label{APPEND_EXAMPLE}

Here we give a few simple examples for the GF of an infinite (or
semi-infinite) system containing a single scatterer. As the first example, a
semi-infinite lead (with unit cell Hamiltonian $\mathbf{h}$, hopping
$\mathbf{t}$, and propagation matrices $\mathbf{P}_{\pm}$) consisting of the
unit cells $m\leq0$ can be regarded as a single-unit-cell scatterer at $m=0$
connected to a semi-infinite left lead. Then the conversion matrix of this
scatterer is
\begin{equation}
\mathbb{G}^{(\mathrm{L})}=(z-\mathbf{h}-\mathbf{t}^{\dagger}\mathbf{P}%
_{-}^{-1})^{-1}=(\mathbf{tP}_{-})^{-1}, \label{GL}%
\end{equation}
where we have used Eq. (\ref{EQUALITY}) in the second step. The GF of the
entire system is%
\begin{equation}
\mathbf{G}_{m,m_{0}}^{(\mathrm{L})}=\mathbf{g}_{m,m_{0}}+\mathbf{P}_{-}%
^{m}(\mathbb{G}^{(\mathrm{L})}\mathbf{g}^{-1}-\mathbf{I})\mathbf{g}_{0,m_{0}%
}\mathbf{.} \label{GL_MM0}%
\end{equation}
Taking $m=m_{0}=0$ gives $\mathbf{G}_{0,0}^{(\mathrm{L})}=\mathbb{G}%
^{(\mathrm{L})}$. By using Eqs. (\ref{GL}) and (\ref{G00}), we have
$\mathbb{G}^{(\mathrm{L})}\mathbf{g}^{-1}-\mathbf{I}=-\mathbf{P}_{-}%
^{-1}\mathbf{P}_{+}$ and hence recover the results by Sanvito \textit{et al}.
\cite{SanvitoPRB1999}: $\mathbf{G}_{m,m_{0}}^{(\mathrm{L})}=\mathbf{g}%
_{m,m_{0}}-\mathbf{P}_{-}^{m-1}\mathbf{P}_{+}^{1-m_{0}}\mathbf{g}$.

As the second example, a semi-infinite lead consisting of the unit cells
$m\geq0$ can be regarded as a single-unit-cell scatterer at $m=0$ connected to
a semi-infinite right lead. Then the conversion matrix of this scatterer is%
\begin{equation}
\mathbb{G}^{(\mathrm{R})}=(z-\mathbf{h}-\mathbf{tP}_{+})^{-1}=(\mathbf{t}%
^{\dagger}\mathbf{P}_{+}^{-1})^{-1}.
\end{equation}
The GF of the entire system is%
\begin{equation}
\mathbf{G}_{m,m_{0}}^{(\mathrm{R})}=\mathbf{g}_{m,m_{0}}+\mathbf{P}_{+}%
^{m}(\mathbb{G}^{(\mathrm{R})}\mathbf{g}^{-1}-\mathbf{I})\mathbf{g}_{0,m_{0}},
\end{equation}
where $\mathbb{G}^{(\mathrm{R})}\mathbf{g}^{-1}-\mathbf{I}=-\mathbf{P}%
_{+}\mathbf{P}_{-}^{-1}$. Taking $m=m_{0}=0$ gives $\mathbf{G}_{0,0}%
^{(\mathrm{R})}=\mathbb{G}^{(\mathrm{R})}$.

The third example is an interface at $m=0$ (with unit cell Hamiltonian
$\mathbf{H}_{0,0}$) connected to two semi-infinite leads $L$ and $R$. In this
case the conversion matrix of the interface is
\begin{equation}
\mathbb{G}^{(\mathrm{I})}=(z-\mathbf{H}_{0,0}-\mathbf{t}_{L}^{\dagger
}(\mathbf{P}_{-}^{(L)})^{-1}-\mathbf{t}_{R}\mathbf{P}_{+}^{(R)})^{-1},
\end{equation}
where $\mathbf{t}_{L}$ ($\mathbf{t}_{R}$) is the nearest-neighbor hopping in
the left (right) lead. The GFs of the entire system are given by%
\begin{align}
\mathbf{G}_{m\geq0,m_{0}\leq0}  &  =(\mathbf{P}_{+}^{(R)})^{m}\mathbb{G}%
^{(\mathrm{I})}(\mathbf{g}^{(L)})^{-1}\mathbf{g}_{0,m_{0}}^{(L)},\\
\mathbf{G}_{m\leq0,m_{0}\geq0}  &  =(\mathbf{P}_{-}^{(L)})^{m}\mathbb{G}%
^{(\mathrm{I})}(\mathbf{g}^{(R)})^{-1}\mathbf{g}_{0,m_{0}}^{(R)},\\
\mathbf{G}_{m\leq0,m_{0}\leq0}  &  =\mathbf{g}_{m,m_{0}}^{(L)}+(\mathbf{P}%
_{-}^{(L)})^{m}[\mathbb{G}^{(\mathrm{I})}(\mathbf{g}^{(L)})^{-1}%
-\mathbf{I}]\mathbf{g}_{0,m_{0}}^{(L)},\\
\mathbf{G}_{m\geq0,m_{0}\geq0}  &  =\mathbf{g}_{m,m_{0}}^{(R)}+(\mathbf{P}%
_{+}^{(R)})^{m}[\mathbb{G}^{(\mathrm{I})}(\mathbf{g}^{(R)})^{-1}%
-\mathbf{I}]\mathbf{g}_{0,m_{0}}^{(R)}.
\end{align}


\end{document}